\documentclass{aa}  

\usepackage{graphicx}
\usepackage{txfonts}
\usepackage[colorlinks=true,citecolor=blue]{hyperref}
\begin{document}

   \title{Ground-based detection of the near-infrared emission from  
          the dayside of \object{WASP-5b}
          \thanks{Based on observations collected with the Gamma 
          Ray Burst Optical and Near-Infrared Detector (GROND) at the MPG/ESO 2.2-meter
          telescope at La Silla Observatory, Chile. Programme 087.A-9006 (PI: Chen).}
          \fnmsep\thanks{Photometric time series are only available in electronic 
                         form at the CDS via anonymous ftp to cdsarc.u-strasbg.fr 
                         (130.79.128.5) or via http://cdsweb.u-strasbg.fr/cgi-bin/qcat?J/A+A/}}


   \author{G. Chen\inst{1,2,3}
          \and
           R. van Boekel\inst{3}
          \and
           N. Madhusudhan\inst{4}
          \and
           H. Wang\inst{1}
          \and
           N. Nikolov\inst{3,5}
          \and
           U. Seemann\inst{6}
          \and
           Th. Henning\inst{3}
          }

   \institute{Purple Mountain Observatory, \& Key Laboratory for Radio Astronomy, Chinese 
             Academy of Sciences, 2 West Beijing Road, Nanjing 210008, PR China\\
              \email{guochen@pmo.ac.cn}
         \and
             University of Chinese Academy of Sciences, No.19A Yuquan Road, 
             100049 Beijing, PR China
         \and
             Max Planck Institute for Astronomy, K\"onigstuhl 17, 69117 Heidelberg, Germany
         \and
             Institute of Astronomy, University of Cambridge, Madingley Road, Cambridge, CB3 0HA, UK
         \and
             Astrophysics Group, University of Exeter, Stocker Road, EX4 4QL, Exeter, UK
         \and
             Institut f\"ur Astrophysik, Friedrich-Hund-Platz 1, 37077 G\"ottingen, Germany
             }

   \date{Received 30 April 2013; accepted 27 February 2014}

   \titlerunning{Near-infrared emission from the dayside of \object{WASP-5b}}
   \authorrunning{G. Chen et al.}

 
  \abstract
   {Observations of secondary eclipses of hot Jupiters allow one to measure 
   the dayside thermal emission from the planets' atmospheres. The combination of 
   ground-based near-infrared observations and space-based observations at 
   longer wavelengths constrains the atmospheric temperature structure and 
   chemical composition.}
   {This work aims at detecting the thermal emission of \object{WASP-5b}, a highly 
   irradiated dense hot Jupiter orbiting a G4V star every 1.6 days, in the $J$, 
   $H$ and $K$ near-infrared photometric bands. The spectral energy distribution 
   is used to constrain the temperature-pressure profile and to study the energy 
   budget of \object{WASP-5b}.}
   {We observed two secondary-eclipse events of \object{WASP-5b} in the $J$, $H$, 
   $K$ bands simultaneously using the GROND instrument on the MPG/ESO 2.2 meter 
   telescope. The telescope was in nodding mode for the first observation 
   and in staring mode for the second observation. The occultation light 
   curves were modeled to obtain the flux ratios in each band, which 
   were then compared with atmospheric models.}
   {Thermal emission of \object{WASP-5b} is detected in the $J$ and $K$ bands in 
   staring mode. The retrieved planet-to-star flux ratios are $0.168^{+0.050}_{-0.052}$\% 
   in the $J$ band and $0.269\pm0.062$\% in the $K$ band, corresponding to brightness 
   temperatures of $2996^{+212}_{-261}$\,K and $2890^{+246}_{-269}$\,K, respectively. 
   No thermal emission is detected in the $H$ band, with a 3-$\sigma$ upper limit of 
   0.166\% on the planet-to-star flux ratio, corresponding to a maximum temperature 
   of 2779~K. On the whole, our $J$, $H$, $K$ results can be explained by a roughly 
   isothermal temperature profile of $\sim$2700~K in the deep layers of the planetary 
   dayside atmosphere that are probed at these wavelengths. Together with {\it Spitzer} 
   observations, which probe higher layers that are found to be at $\sim$1900~K, a 
   temperature inversion is ruled out in the range of pressures probed by the combined 
   data set. While an oxygen-rich model is unable to explain all the data, a carbon-rich 
   model provides a reasonable fit but violates energy balance. The nodding-mode 
   observation was not used for the analysis because of unremovable systematics. Our 
   experience reconfirms that of previous authors: staring-mode observations are 
   better suited for exoplanet observations than nodding-mode observations.}
   {}


   \maketitle
\section{Introduction}\label{sec:intro}

   Currently, the most fruitful results on the characterization of exoplanetary 
   atmospheres come from transiting planets. Since the first transiting planet 
   \object{HD 209458b} was discovered in 1999 \citep{2000ApJ...529L..45C}, 
   more than 400 are confirmed\footnote{http://exoplanet.eu/}. The orbital 
   parameters of these planets are well constrained when transit observations 
   were combined with radial velocity measurements. Precise planetary parameters 
   such as mass and radius can be determined as well, which leads to a preliminary 
   view of the internal structure of a planet, and thus to constrain the formation 
   and evolutionary history of the planet \citep{2005AREPS..33..493G,2007ApJ...659.1661F}. 
   Furthermore, transiting planets provide unprecedented opportunities to probe 
   their atmospheres, not only from wavelength-dependent effective radius variations 
   determined from the transit \citep[e.g.:][]{2002ApJ...568..377C}, but also from 
   differential planetary photon measurements from occultation 
   \citep[e.g.:][]{2005Natur.434..740D}. In the latter case, the planet passes 
   behind the star, which leaves us only stellar emission during a total eclipse.
   
   As a subset of transiting planets that are exposed to high irradiation in close 
   orbits around their host stars, hot Jupiters are the most favorable targets for 
   thermal emission detection through secondary-eclipse observation. Their close 
   orbits translate into a high occultation probability and frequency, while their high 
   temperatures and large sizes make the planet-to-star flux ratio favorable. The 
   first thermal emission detections of hot Jupiters have been achieved with the 
   {\it Spitzer Space Telescope} \citep{2005Natur.434..740D,2005ApJ...626..523C}, 
   which operates in the mid-infrared (MIR) wavelength range. Since then, a flood of such 
   detections have been made with {\it Spitzer} observations, resulting in better 
   knowledge of the chemical composition and thermal structure of the planetary 
   atmosphere. Compared with the MIR, the near-infrared (NIR) wavelength range 
   covers the peak of the spectral energy distribution (SED) of a planet and probes 
   deeper into the atmosphere, therefore it can be used to constrain the atmosphere's 
   temperature structure and energy budget. While the {\it Hubble Space Telescope} 
   has contributed much to the NIR observation on planetary secondary 
   eclipses \citep[e.g.:][]{2009ApJ...690L.114S,2009ApJ...704.1616S}, now more 
   observations with high precision are starting to come from ground-based 
   mid-to-large aperture telescopes thanks to the atmospheric window in the NIR 
   \citep[e.g.:][]{2010ApJ...717.1084C,2010ApJ...718..920C,2011AJ....141...30C,
   2011A&A...530A...5C,2012A&A...542A...4G}. As shown for example by \citet{2012ApJ...758...36M}, 
   these ground-based NIR measurements play a crucial role in determining 
   the C/O ratio when combined with measurements from {\it Spitzer} observations. 
   
   \object{WASP-5b} was first detected by \citet{2008MNRAS.387L...4A} as a hot Jupiter 
   orbiting a 12.3 mag G4V type star every 1.628 days. Its mass and radius are derived 
   to be 1.58 and 1.09 times of the Jovian values, respectively, which places it among 
   the relatively dense hot Jupiters. Several follow-up transit observations have 
   refined its density to be nearly the same as our Jupiter \citep{2009MNRAS.396.1023S,
   2011PASJ...63..287F}. Its host star has a slightly supersolar metallicity 
   ([Fe/H]=+0.09$\pm$0.09), according to the high-resolution VLT/UVES spectroscopy of 
   \citet{2009A&A...496..259G}. The planetary orbit might have a marginally nonezero 
   eccentricity based on the joint analysis of RV and photometric measurements 
   \citep{2009A&A...496..259G,2010A&A...524A..25T,2012MNRAS.422.3151H}. 
   \citet{2010A&A...524A..25T} studied the Rossiter-McLaughlin effect in the 
   \object{WASP-5} system and found a sky-projected spin-orbit angle compatible 
   with zero ($\lambda$=12.1$^{+10.0\circ}_{-8.0}$), indicating an orbit aligned 
   with the stellar rotation axis. Furthermore, several studies focused on the potential 
   transit-timing variations (TTVs) of this system. \citet{2009A&A...496..259G} 
   first noticed that a linear fit cannot explain the transit ephemeris very 
   well, which was later suspected to be caused by the poor quality of one timing 
   \citep{2009MNRAS.396.1023S}. \citet{2011PASJ...63..287F} studied its TTVs in 
   detail with an additional seven new transit observations and calculated a TTV 
   rms of 68\,s, only marginally larger than their mean timing uncertainty of 
   41\,s. \citet{2012ApJ...748...22H} revisited this TTV signal by combining 
   their nine new epochs and suggested that this TTV might be introduced by 
   data uncertainties and systematics and not by gravitational perturbations. 

   From these intensive previous studies, \object{WASP-5b} has become an intriguing 
   target for atmospheric characterization. It is not bloated, although it receives 
   a relatively high irradiation of $\sim$2.1$\times$10$^9$~erg\,s$^{-1}$\,cm$^{-2}$ 
   from its 5700~K \citep{2009A&A...496..259G} host star \citep[assuming a scaled major-axis 
   $a/R_*$=5.37,][]{2011PASJ...63..287F}, which would place it in the pM class in the 
   scheme proposed by \citet{2008ApJ...678.1419F}. Its proximity to the host star results 
   in an equilibrium temperature of 1739~K assuming zero albedo and isotropic redistribution 
   of heat across the whole planet, which could be as high as 2223~K in the extreme case of 
   zero heat-redistribution. Its \ion{Ca}{II} H and K line strength 
   \citep[$\log R'_{\rm{HK}}$=$-$4.72$\pm$0.07,][]{2010A&A...524A..25T} suggests that the 
   activity of the host star might prevent it from having an inverted atmosphere, given the 
   correlation proposed by \citet{2010ApJ...720.1569K}. Recently, \citet{2013ApJ...773..124B} 
   reported thermal detections from the Warm {\it Spitzer} mission, suggesting a weak thermal 
   inversion or no inversion at all, with poor day-to-night energy redistribution.
   
   In this paper, we present the first ground-based detections of thermal emission 
   from the atmosphere of \object{WASP-5b} in the $J$ and $K$ bands through observations 
   of secondary eclipse. Section~\ref{sec:obs} describes our observations of two 
   secondary-eclipse events and the process of data reduction. Section~\ref{sec:lc} 
   summarizes the approaches that we employed to remove the systematics and to optimally 
   retrieve the flux ratios. In Sect.~\ref{sec:discuss}, we discuss remaining systematic 
   uncertainties and orbital eccentricity, and we also offer explanations for the thermal 
   emission of \object{WASP-5b} with planetary atmosphere models. Finally, we conclude 
   in Sect.~\ref{sec:con}.

\section{Observations and data reduction}\label{sec:obs}

   We observed two secondary-eclipse events of \object{WASP-5b} with the GROND 
   instrument mounted on the MPG/ESO 2.2 meter telescope at La Silla, Chile. 
   This imaging instrument was designed primarily for the simultaneous observation 
   of gamma-ray burst afterglows and other transients in seven 
   filters: the Sloan $g'$, $r'$, $i'$, $z'$ and the NIR $J$, $H$, $K$ 
   \citep{2008PASP..120..405G}. Dichroics are used to split the incident light 
   into seven optical and NIR channels. Photons of the four optical channels are 
   recorded by backside-illuminated $2048\times2048$ E2V CCDs and stored in FITS 
   files with four extensions. Photons of the three NIR channels are 
   recorded by Rockwell HAWAII-1 arrays ($1024\times1024$) and stored in a single 
   FITS file with a size of $3072\times1024$. The optical arm has a field of view 
   (FOV) of $5.4\times5.4$\,arcmin$^2$ with a pixel scale of $0\farcs158$, while 
   the NIR arm has an FOV of $10\times10$\,arcmin$^2$ with a pixel scale of $0\farcs60$. 
   The guiding system employs a camera placed outside the main GROND vessel, $23'$ 
   south of the scientific FOV, which has a crucial impact on the choice of science 
   pointing especially in the case of defocused observations.\footnote{To avoid 
   poor guiding, we did not employ the defocusing technique.} The capability of 
   simultaneous optical-to-NIR multiband observation makes GROND a potentially 
   good instrument for transit and occultation observations. For secondary-eclipse 
   observations, the optical arm provides the opportunity to detect scattered 
   light in favorable cases, while the NIR arm allows one to construct an SED 
   for the thermal emission of a planetary atmosphere. In both of our observations, 
   we only used the NIR arm (i.e. \object{WASP-5} was not in the optical FOV) 
   to include as many potential reference stars as possible in the NIR FOV. 
   
\begin{figure}
     \centering
     \includegraphics[width=\hsize]{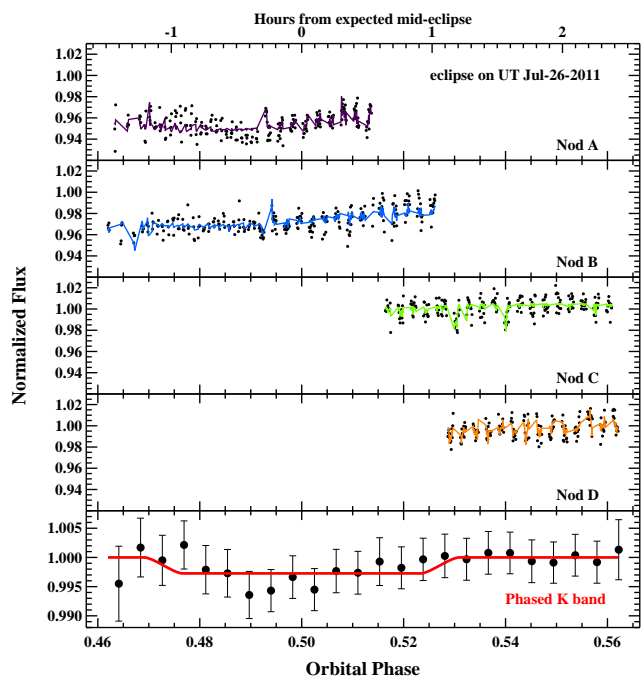}
     \caption{\footnotesize $K$-band occultation light-curve of \object{WASP-5} 
            observed on UT July 26 2011 (in nodding mode). As described in the text, 
            instrumental crashes led to four groups of nods, none of which covered 
            the whole eclipse. \textit{The top four panels} show the raw light-curve for 
            each nod, overplotted with the best-fit model. \textit{Bottom panel} shows 
            the phase-folded $K$-band light curve, which has been corrected for baseline 
            trend and binned every 10 minutes for display purposes. \label{fig:nod}}
   \end{figure}

   The first secondary-eclipse event was observed continuously for four hours on UT 
   July 26 2011, from 04:13 to 08:16. The observation was performed in an ABAB nodding 
   pattern. Four exposures were taken on each nodding position during the first one 
   third of the observing time, and 12 exposures each were taken in the remaining time. 
   Each exposure was composed of two integrations of 3 seconds each (DIT=3~s), which were 
   averaged together. However, the actual nodding pattern was far more complicated. The 
   telescope operation GUI software crashed several times, and a new ABAB pattern was 
   re-started on each crashed position. The resulting time series of each band was full 
   of red noise, which is difficult to correct since systematic effects affect the 
   recorded signals differently depending on location on the detector. Time series of 
   each location did not cover the whole occultation duration, which makes the systematic 
   correction problem even worse. Therefore we decided to discard this dataset in our 
   further analysis. Only the result in the $K$ band is shown for comparison in 
   Sect.~\ref{sec:lc} and Fig.~\ref{fig:nod}.

   The second event of the secondary eclipse was observed continuously for 4.6 hours on UT 
   September 8 2011, from 02:41 to 07:17 in staring mode. Before and after the science time 
   series, the sky around the scientific FOV was measured using a 20-position dither pattern, 
   which was used to construct the sky emission model in the subsequent reduction. During the 
   science observation, four integrations of 3 seconds each were averaged into one exposure, 
   resulting in 707 frames recorded and a duty cycle of $\sim$53\%. The peak count level 
   of the target star is well below the saturation level. The airmass started at 1.22, 
   decreased to 1.02, and rose to 1.11 in the end; the seeing was unstable during the eclipse, 
   ranging from $1''$ to $3''$ as measured from the point spread functions (PSFs) of the stars. 
   The moon was illuminated around 82\%, and had a minimum distance of 55$^{\circ}$ to 
   \object{WASP-5} at the end of the observation. In the following text, we always refer to 
   this second dataset unless specified otherwise.
   
   We reduced the acquired data with our IDL\footnote{IDL is an acronym for Interactive Data 
   Language, for details we refer to http://www.exelisvis.com/idl/} pipeline in a standard 
   way, which mainly makes use of NASA IDL Astronomy User's Library\footnote{See 
   http://idlastro.gsfc.nasa.gov/}. General image calibration steps\footnote{In principle, 
   NIR data need nonlinearity correction. However, we did not include the correction in our 
   final calibration steps to avoid introducing additional noise 
   \citep[similar to e.g.][]{2013ApJ...771..108B}. According to our experiment, the derived 
   eclipse depths did not change significantly ($\ll1\sigma$) when the data were reduced with 
   or without nonlinearity correction.} include dark subtraction, read-out pattern removal, 
   flat division, and sky subtraction. We made DARK master files by median-combining 20 individual 
   dark-current measurements and subtracted them from all the raw images. To correct the 
   electronic odd-even readout pattern along the X-axis, each dark-subtracted image was smoothed 
   with a boxcar median filter and compared with the unsmoothed one. The amplitudes of readout 
   patterns were obtained from the resulting difference image and were corrected in the unsmoothed 
   dark-subtracted image. Finally, SKYFLAT master files were generated by median-combining 48 
   individual twilight sky flat measurements, which first had the star masked out and were then 
   normalized and combined. The dark- and pattern-corrected images were divided by these SKYFLAT 
   files for flat-field correction. 
   
   To eliminate the sky contribution in our staring-mode data, we constructed a sky emission 
   model for each science image using the 20-position dithering sky measurements. These sky 
   images were star-masked and normalized and then dark- and flat-calibrated in the same way
   as the science images. We median-combined on the sky stack images to generate 
   basic sky emission models. The pre- and post-science sky models were scaled to the background 
   level of each science image. A final sky model was created by combining the pre- and 
   post-science sky models while taking the inverse square of the fitted $\chi^2$ as the weight.  
   The final sky model was later subtracted from the corresponding science image. Due to the long 
   time-scale of our observation, the sky is expected to be variable. Thus this sky correction 
   is only a first-order correction. Nevertheless, it results in light curves of slightly 
   better precision than the approach without sky subtraction, according to our experiment.
   
   We performed aperture photometry on the calibrated images with the IDL DAOPHOT package. We first 
   determined the locations of \object{WASP-5} as well as several nearby comparison stars of 
   similar brightness using IDL/FIND, which calculates the centroids by fitting Gaussians to 
   the marginal $x$ and $y$ distributions. The FWHMs for each star, which were used to indicate 
   the seeing during our observation, were calculated in a similar way. We carefully chose 
   the best comparison-star ensemble to normalize the \object{WASP-5} time series as follows: 
   various combinations of comparison stars were tried. For each ensemble, time series of chosen 
   comparison stars (as well as \object{WASP-5}) were individually normalized by the median of 
   their out-of-eclipse flux levels, and then weighted-combined according to the inverse square 
   of uncertainties. The ensemble that made the normalized \object{WASP-5} light curve show 
   the smallest scatter was considered as the optimal reference. We also experimented to find 
   the best photometric results by placing 30 apertures on each star in a step of 0.5 pixel, 
   each aperture again with 10 annuli of different sizes in a step of 1 pixel. The aperture and 
   annulus that made reference-corrected \object{WASP-5} light curve behave with the smallest 
   scatter was chosen as the optimal aperture setting. As a result, we used six comparison stars 
   for the $J$ band, three for the $H$ band, and four for the $K$ band. The aperture settings 
   for the $J$, $H$, $K$ bands are (6.5, 13.5-22.0) pixels, (6.0, 6.0-19.0) pixels, (5.0, 7.0-19.0) 
   pixels in the format of (aperture size, sky annulus inner/outer sizes), respectively. 
   
   Finally, we extracted the time stamp stored in the header of the FITS file. The default 
   time stamp was the starting UTC time of each frame. We took into account the readout 
   time and the arm-waiting time\footnote{The optical and NIR arms of GROND are not operated 
   independently.} to make the final time stamp centered on the central point of each total 
   integration. We converted this UTC time stamp into Barycentric Julian Date in the 
   Barycentric Dynamical Time standard (BJD$_{\rm{TDB}}$) using the IDL procedure written 
   by \citet{2010PASP..122..935E}.
   
   \begin{figure*}
     \centering
     \includegraphics[width=0.4\hsize]{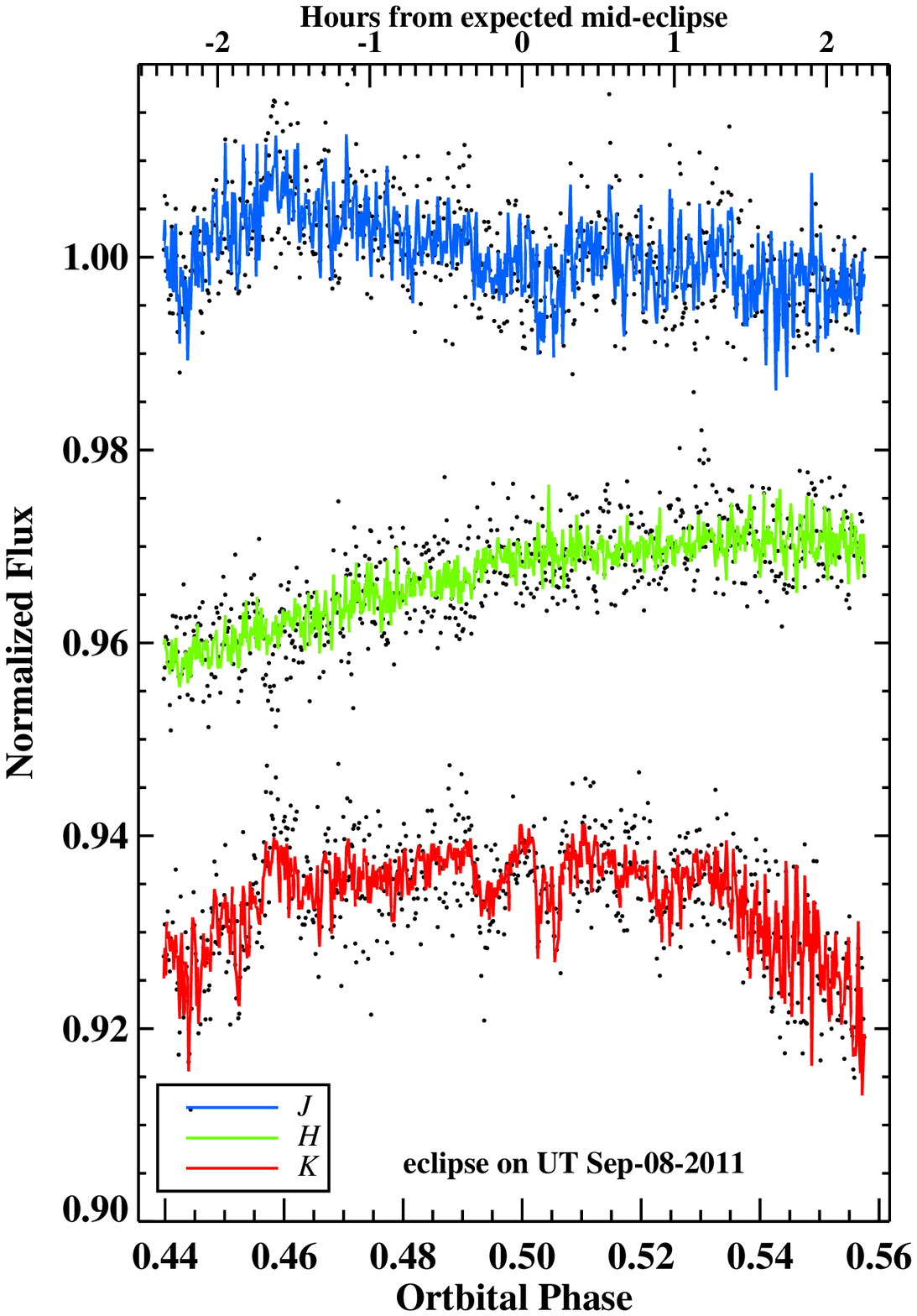}
     \includegraphics[width=0.4\hsize]{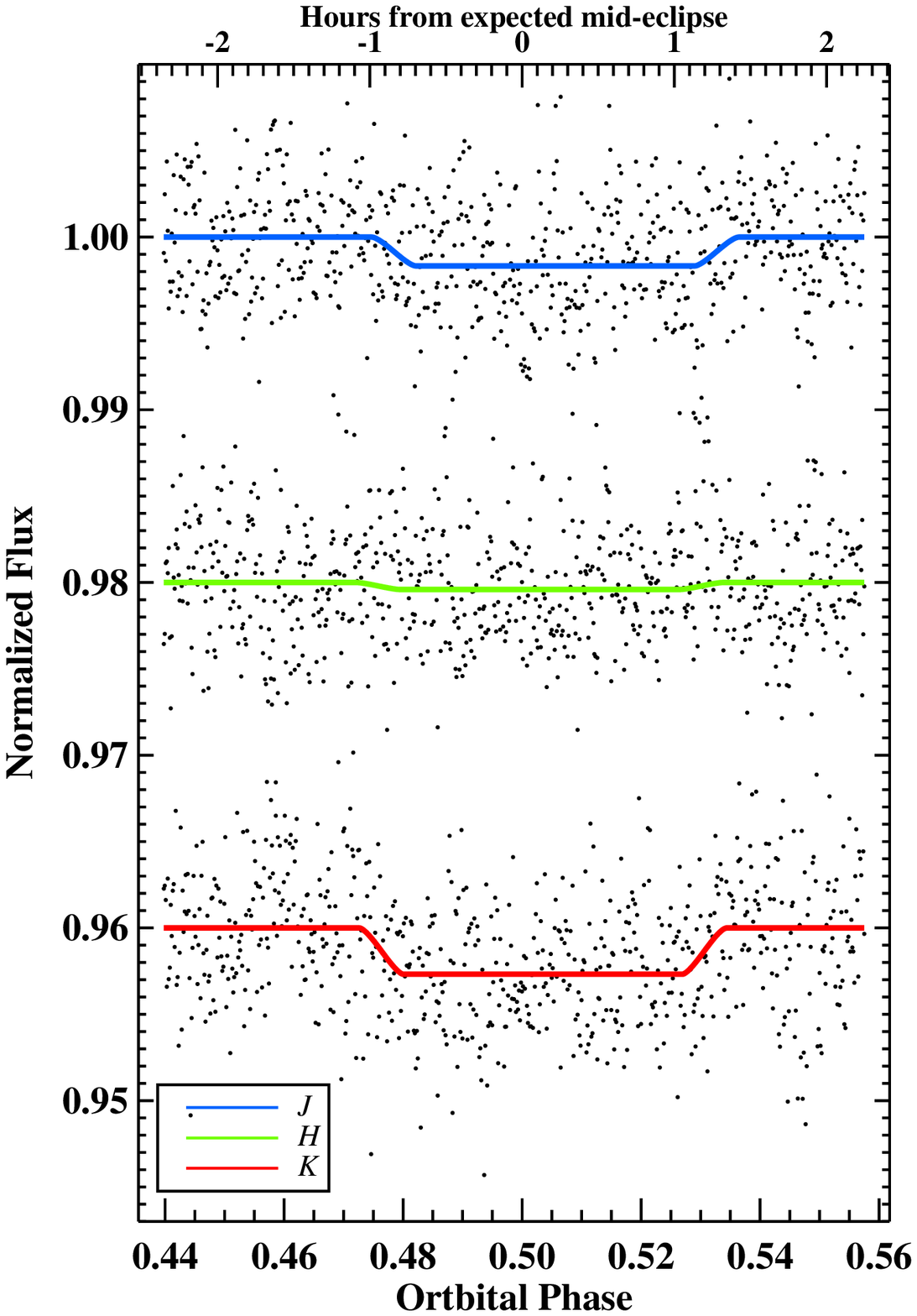}\\
     \includegraphics[width=0.4\hsize]{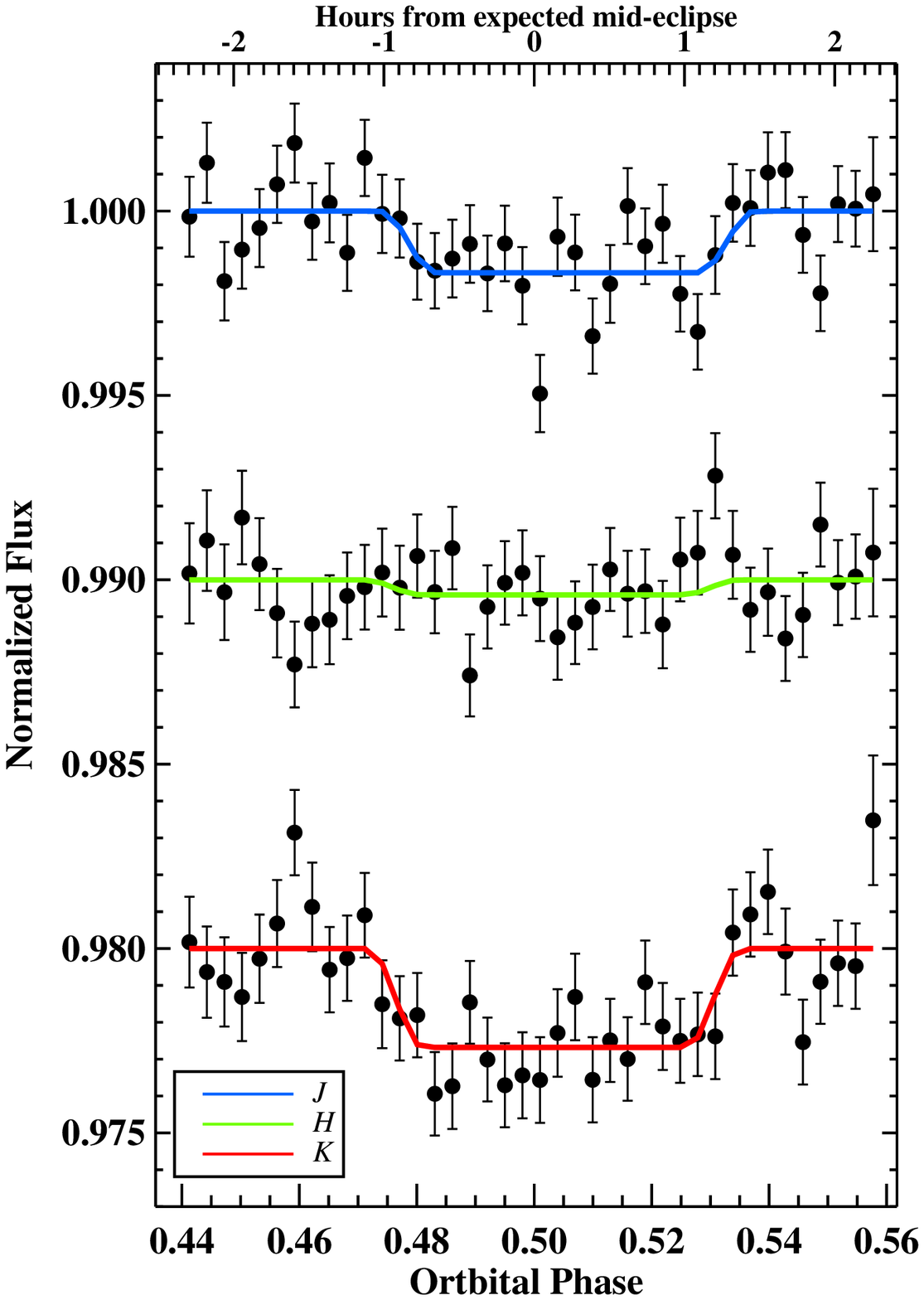}
     \includegraphics[width=0.4\hsize]{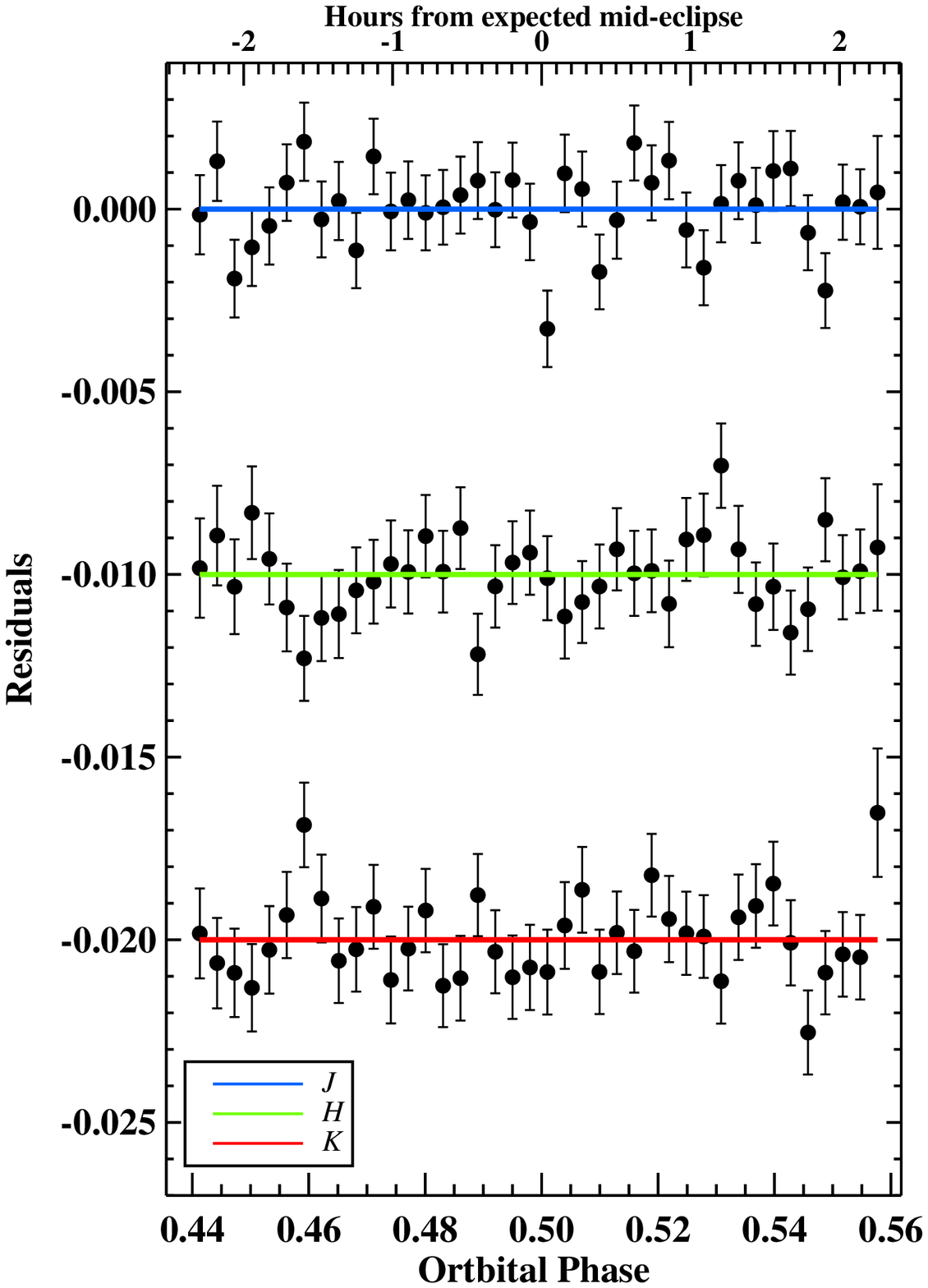}
     \caption{\footnotesize Near-infrared occultation light-curves of \object{WASP-5} 
            observed on UT Sep 08 2011 (in staring mode). Each panel from top to bottom 
            shows for the $J$, $H$ and $K$ bands. \textit{Top left:} raw light-curves
            overplotted with the best-fit models, which consist of a theoretical light-curve
            multiplied with a decorrelation function. \textit{Top right:} light curves 
            after baseline correction. \textit{Bottom left:} light curves binned every 
            seven minutes for display purposes. \textit{Bottom right:} residuals of 
            light-curve fitting.\label{fig:occ}}
   \end{figure*}

   \begin{table*}
     \caption{Results of the MCMC analysis on the secondary eclipse (Sep-08-2011) of \object{WASP-5b}\label{tab:result}}
     \centering
     \begin{tabular}{lcccc}
     \hline\hline
     Parameter & Units & $J$ band & $H$ band & $K$ band\\
     \hline
     $T_{\rm{mid,occ}}$ & BJD$_{\rm{TDB}}-$2450000 & 5812.7249 $^{+0.0032}_{-0.0031}$ 
                                                   & 5812.7207 $^{+0.0031}_{-0.0031}$ 
                                                   & 5812.7215 $^{+0.0029}_{-0.0030}$\\
     $\phi_{\rm{mid,occ}}$\tablefootmark{a}  & ... & 0.5054 $^{+0.0020}_{-0.0019}$ 
                                                   & 0.5028 $^{+0.0019}_{-0.0019}$ 
                                                   & 0.5033 $^{+0.0018}_{-0.0018}$\\
     $T_{\rm{offset}}$\tablefootmark{a}  & minutes & 12.6 $^{+4.7}_{-4.5}$ 
                                                   & 6.5 $^{+4.5}_{-4.5}$ 
                                                   & 7.7 $^{+4.2}_{-4.3}$\\ 
     $F_p/F_*$ & \% & 0.168 $^{+0.050}_{-0.052}$ 
                    & 0.041 $^{+0.039}_{-0.028}$, <0.166 (3$\sigma$)
                    & 0.269 $^{+0.062}_{-0.062}$ \\
     $T_B$ & K & 2996 $^{+212}_{-261}$ 
               & <2779 (3$\sigma$)
               & 2890 $^{+246}_{-269}$\\
     $T_{58}$ & days & 0.1001 $^{+0.0063}_{-0.0070}$
                     & ... 
                     & 0.1026 $^{+0.0054}_{-0.0058}$\\ 
     $e\cos\omega$ & ... & 0.0085 $^{+0.0030}_{-0.0029}$
                         & ...
                         & 0.0052 $^{+0.0028}_{-0.0029}$\\
     $e\sin\omega$ & ... & -0.002 $^{+0.036}_{-0.044}$
                         & ...
                         & 0.013 $^{+0.030}_{-0.035}$\\
     $e$ & ...           & 0.027 $^{+0.032}_{-0.015}$
                         & ...
                         & 0.025 $^{+0.024}_{-0.015}$\\
     $|\omega|$ & $^{\circ}$ & 72 $^{+11}_{-32}$
                             & ...
                             & 79 $^{+7}_{-24}$\\
     \hline
        Baseline & ... & Equation (\ref{equ:bfj}) & Equation (\ref{equ:bfh}) & Equation (\ref{equ:bfk})\\
        $c_{0}$ & ...  & 1.00145 $^{+0.00031}_{-0.00033}$ & 1.00343 $^{+0.00022}_{-0.00015}$
                       & 1.00916 $^{+0.00052}_{-0.00053}$\\
        $c_{1}$ & ...  & 0.01158 $^{+0.00012}_{-0.00012}$ & 0.004669 $^{+0.000053}_{-0.000077}$
                       & -0.00173 $^{+0.00024}_{-0.00022}$\\
        $c_{2}$ & ...  & 0.00808 $^{+0.00052}_{-0.00049}$ & -0.009248 $^{+0.000027}_{-0.000048}$
                       & -0.00450 $^{+0.00022}_{-0.00023}$\\
        $c_{3}$ & ...  & -0.00233 $^{+0.00027}_{-0.00028}$ & 0.00067 $^{+0.00015}_{-0.00010}$
                       & -0.001958 $^{+0.00010}_{-0.00011}$\\
        $c_{4}$ & ...  & 0.00715 $^{+0.00028}_{-0.00027}$ & 0.001877 $^{+0.000044}_{-0.000092}$
                       & -0.00864 $^{+0.00012}_{-0.00011}$\\
        $c_{5}$ & ...  & -0.00760 $^{+0.00015}_{-0.00015}$ & -0.000941 $^{+0.00010}_{-0.00014}$
                       & 0.0265 $^{+0.0046}_{-0.0043}$\\
        $c_{6}$ & ...  & -0.00209 $^{+0.00033}_{-0.00031}$ & ... & ...\\
        $c_{7}$ & ...  & -0.00459 $^{+0.00014}_{-0.00014}$ & ... & ...\\
        $c_{8}$ & ...  & -0.0246 $^{+0.0027}_{-0.0027}$ & ... & ...\\
     \hline
     \end{tabular}
   \tablefoot{
       \tablefoottext{a}{Light travel time ($\sim$27 s) in the system has been corrected.}
             }
   \end{table*}

\section{Light-curve analysis}\label{sec:lc}

   As shown in the top left panel of Fig.~\ref{fig:occ}, the \object{WASP-5} light curves 
   exhibited obvious red noise even after normalization by the composite reference light-curve 
   as described above. Part of this red noise arises from instrumental systematics, 
   such as different star locations on the detector, seeing variation (thus different number 
   of pixels within the volume of the star's FWHM), which can be inferred from the correlation 
   between each parameter and the normalized flux (see Fig.~\ref{fig:app_fig3}--\ref{fig:app_fig4} 
   in the appendix). In the literature, some authors chose to construct a 
   systematics model using out-of-eclipse data and applied these relationships to the 
   whole light curve for correction \citep[e.g.][]{2010ApJ...717.1084C}. This requires that 
   the range of instrumental parameters during in-eclipse is repeatable in the out-of-eclipse 
   data, otherwise it would lead to extrapolation. Since most of our instrumental parameters 
   were not necessarily repeatable between in-eclipse and out-of-eclipse (e.g. slow drift 
   of star location on the detector, variation of seeing), we decided to fit the whole 
   light curve with an analytic occultation model multiplied by a baseline correction model. 
   
   We adopted the \citet{2002ApJ...580L.171M} formulae without limb-darkening as our occultation 
   model. System parameters such as period $P$, planet-to-star radius ratio $R_p/R_*$, inclination 
   $i$, and scaled semi-major axis $a/R_*$ were obtained from \citet{2011PASJ...63..287F} and 
   were fixed in the formulae, while mid-occultation time $T_{\rm{mid}}$ and flux ratio $F_p/F_*$ 
   were set as free parameters. The baseline detrending model was a sum of polynomials of star 
   positions ($x$, $y$), seeings ($s$), airmass ($z$), and time ($t$). We varied the combination 
   of these instrumental and atmospheric terms to generate different baseline models. We searched 
   for the best-fit solutions by minimizing the chi-square: 
   \begin{equation}
     \chi^2=\sum\limits_{i=1}^{N}\frac{[f_i(\mathrm{obs}) - f_i(\mathrm{mod})]^2}{\sigma_{f,i}^2},
   \end{equation}
   in which $f_i(\rm{obs})$ and $\sigma_{f,i}$ are the light-curve data and its uncertainty, while 
   $f_i(\rm{mod})$ is the light-curve model, in the form of 
   \begin{equation}
     f(\mathrm{mod})=E(T_{\mathrm{mid}},F_p/F_*)B(x,y,s,z,t),
   \end{equation}
   
   We experimented with a set of baseline models to find the model that can best remove the 
   instrumental systematics. We calculated the Bayesian information criterion \citep[BIC,][]{Schwarz1978} 
   for the results from different baseline models:
   \begin{equation}
     BIC=\chi^2+k\log(N),
   \end{equation}
   where $k$ is the number of free parameters and $N$ is the number of data points. The baseline 
   model that generated the smallest BIC value was considered as our final choice. With this 
   approach, we used as few free parameters as possible to prevent overinterpreting the baseline 
   function. In our experiments, linear baseline functions in most cases failed to fit the eclipse 
   depth and produced very large BIC values. Among the baseline functions that have BICs similar 
   to that of the chosen one, the measured eclipse depths agree well with each other (see e.g. 
   Fig.~\ref{fig:app_fig1} and \ref{fig:app_fig2}). The final adopted baseline functions are
   \begin{equation}\label{equ:bfj}
     B_J=c_0+c_1x+c_2y+c_3xy+c_4x^2+c_5y^2+c_6s_x+c_7s_y+c_8t,
   \end{equation}
   \begin{equation}\label{equ:bfh}
     B_H=c_0+c_1x+c_2y+c_3xy+c_4y^2+c_5s_x, 
   \end{equation}
   and 
   \begin{equation}\label{equ:bfk}
     B_K=c_0+c_1xy+c_2x^2+c_3y^2+c_4s+c_5t, 
   \end{equation}
   where $s_x$ and $s_y$ refer to the FWHMs of marginal $x$ and $y$ distributions, respectively, 
   while $s$ is their quadratic mean.  
   
   We employed the Markov chain Monte Carlo (MCMC) technique with the Metropolis-Hastings 
   algorithm with Gibbs sampling to determine the posterior probability distribution function 
   (PDF) for each parameter \citep[see e.g.][]{2005AJ....129.1706F,2006ApJ...642..505F}. 
   Following the approach of \citet{2010A&A...511A...3G}, only parameters in the analytic 
   occultation model $E(T_{\rm{mid}},F_p/F_*)$ are perturbed, while the coefficients of 
   baseline function are solved using the singular value decomposition \citep[SVD,][]{press1992} 
   algorithm. At each MCMC step, a jump parameter was randomly selected, and the light curve 
   was divided by the resulting analytic occultation model. The coefficients in the baseline 
   function were then solved by linear least-squares minimization using the SVD. This jump was 
   accepted if the resulting $\chi^2$ is lower than the previous $\chi^2$, or accepted 
   according to the probability $\exp{(-\Delta\chi^2/2)}$ if the resulting $\chi^2$ is 
   higher. We optimized the step scale so that the acceptance rate was $\sim$0.44 before 
   a chain starts \citep{2006ApJ...642..505F}. After running a chain of MCMC, the first 
   10\% links were discarded and the remaining were used to determine the best-fit values 
   and uncertainties of jump parameters (as well as the baseline coefficients). Several 
   chains were run to check that they were well mixed and converged using the 
   \citet{gelman1992} statistics. We adopted the median values of the marginalized 
   distributions as the final parameter values and the 15.865\%/84.135\% values of 
   the distributions as the 1-$\sigma$ lower/upper uncertainties, respectively. 
   
   \begin{figure*}
     \centering
     \includegraphics[width=0.3\hsize]{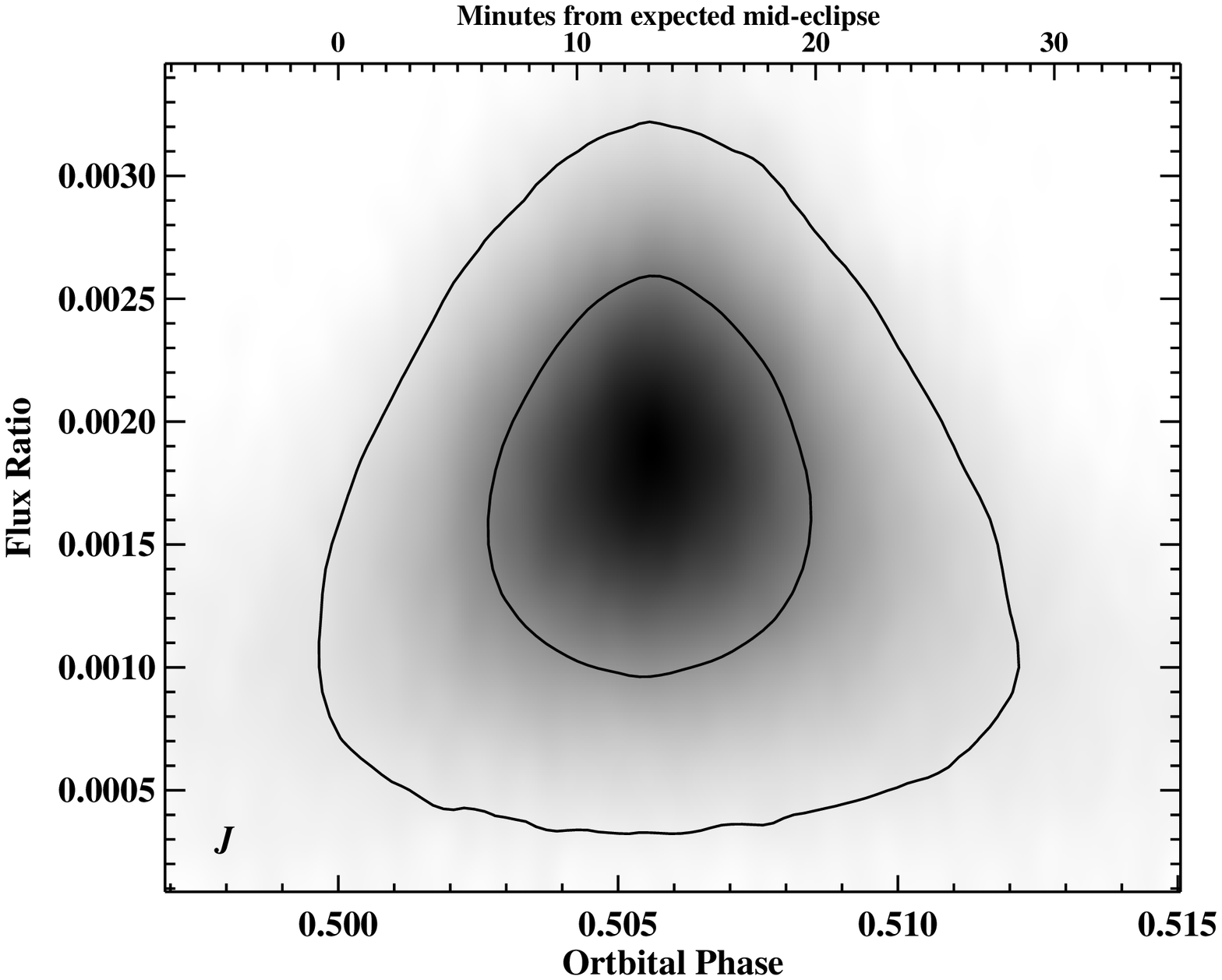}
     \includegraphics[width=0.3\hsize]{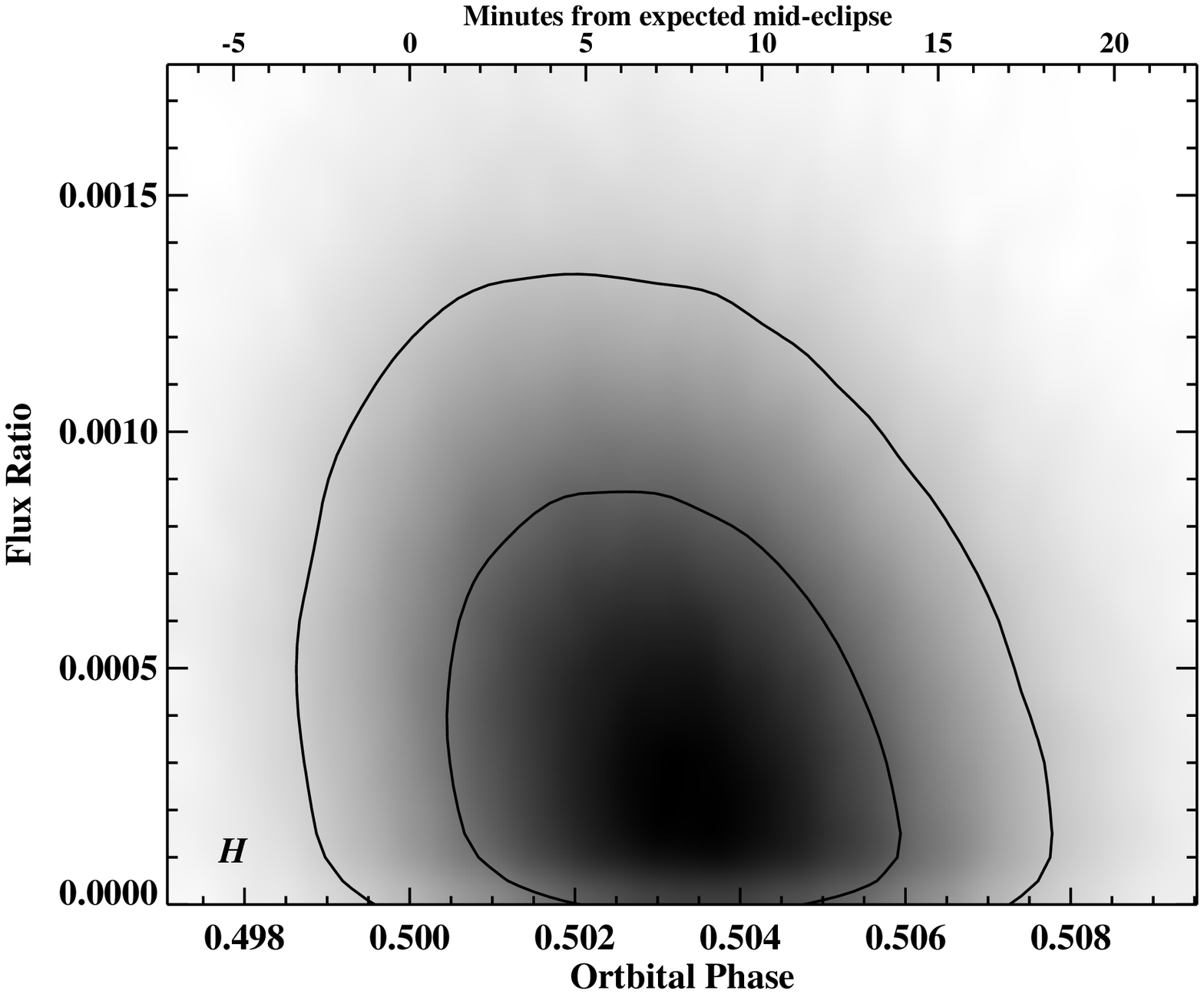}
     \includegraphics[width=0.3\hsize]{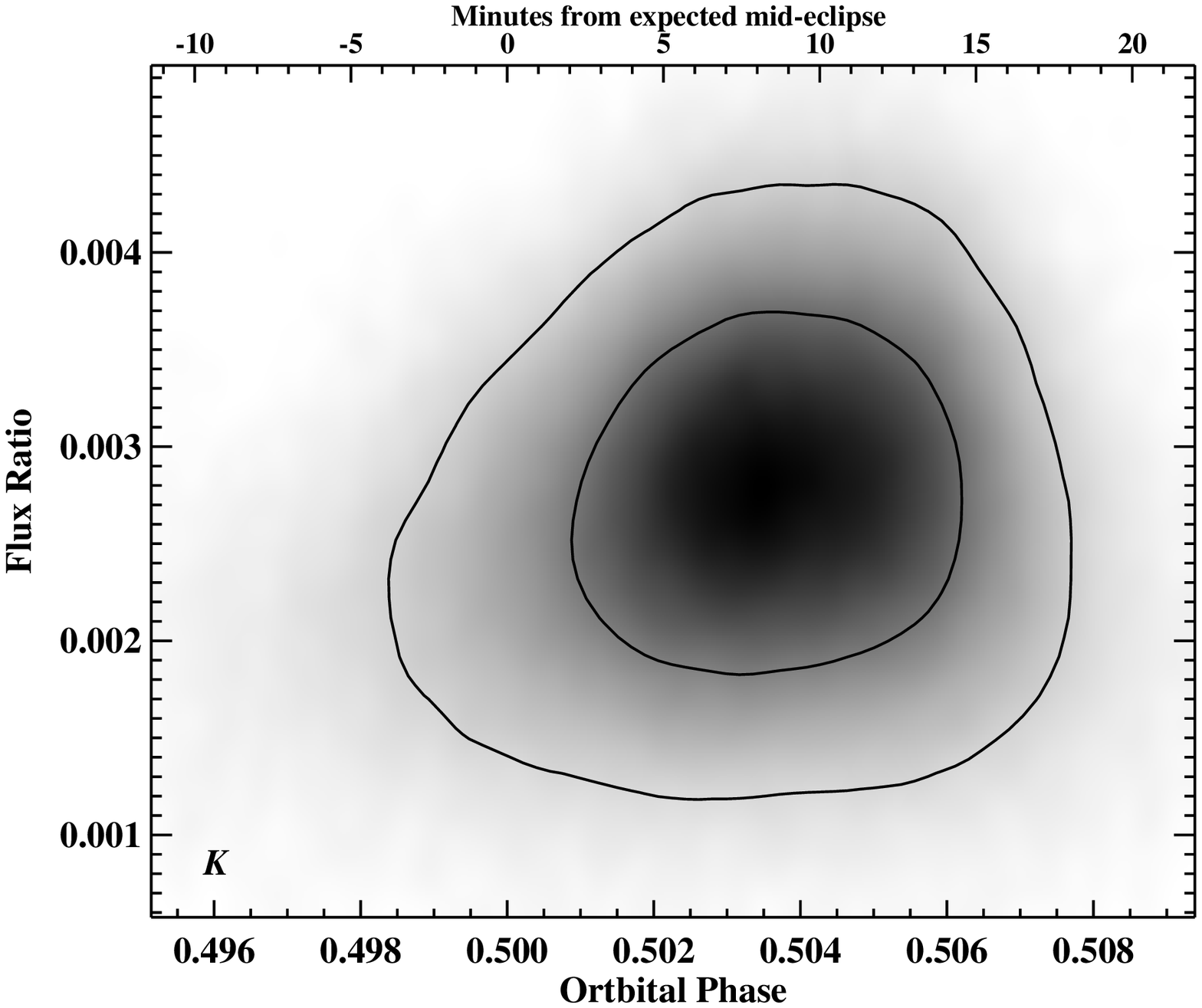}
     \caption{\footnotesize Joint probability distributions between mid-eclipse time and flux ratio 
              from our MCMC analysis for the $J$ (\textit{left}), $H$ (\textit{middle}), and $K$ bands 
              (\textit{right}). The mid-eclipse time has been converted to phase in these plots. The 
              contour lines mark the 68.3\% ($1\sigma$) and 95.5\% ($2\sigma$) confidence regions 
              of the joint posterior distributions, respectively. The gray scale indicates the 
              density of distributions. \label{fig:corr}}
   \end{figure*}
   
   We performed the MCMC-based light-curve modeling in three scenarios. In the first scenario, 
   we tried to find the best-fit mid-occultation times and flux ratios. $T_{\rm{mid}}$ and $F_p/F_*$ 
   were allowed to vary freely for the $J$ and $K$ bands. Since there was no detection in the 
   $H$ band, we chose to adopt the PDF of $T_{\rm{mid}}$ coming from the $K$ band as a Gaussian 
   prior. We first ran a chain of 1\,000\,000 links to find the scaling factors, since the photometric 
   uncertainties might not represent the real uncertainties in the light curves well. We calculated 
   the reduced chi-square ($\chi^2_{\nu}$) for the best-fit model and recorded the scaling factor 
   $\beta_1$=$\sqrt{\chi^2_v}$. We then calculated the standard deviations for the best-fit residuals 
   without binning, also for those with time bins ranging from 10 minutes to the ingress/egress 
   duration of \object{WASP-5b}. The median value of the factor
   \begin{equation}
     \beta_2=\frac{\sigma_N}{\sigma_1}\sqrt{\frac{N(M-1)}{M}}
   \end{equation}
   was recorded as the second scaling factor, where $N$ is the number of individual binned points, 
   $M$ is the number of bins, $\sigma_N$ is the standard deviation of $N$-point binned residuals, 
   and $\sigma_1$ is the un-binned version. These two scaling factors were multiplied with the original 
   uncertainties to account for the under-/overestimated noise. Normally, the first scaling 
   factor would make the fitting to have a reduced chi-square close to 1, while the second would take 
   into account the time-correlated red noise. This approach has been widely applied in transit 
   light-curve modeling \citep[see e.g.:][]{2006MNRAS.373..231P,2008ApJ...683.1076W}. The derived 
   ($\beta_1$, $\beta_2$) for the $J$, $H$, $K$ bands were (1.79, 1.28), (1.25, 1.46), (1.05,1.33), 
   respectively. After rescaling the uncertainties, we ran another five chains of 1\,000\,000 links 
   to finalize the modeling. The $J$-band flux ratio changed from 0.175$\pm$0.021\% to 
   0.168$^{+0.050}_{-0.052}$\%, while the $K$-band flux ratio changed from 0.272$\pm$0.044\% to 
   0.269$\pm$0.062\%. The achieved light-curve quality for the $J$, $H$ and $K$ bands are 1847, 1813 
   and 1777~ppm per two-minute interval in terms of $rms$ of O--C (observed minus calculated) residuals. 
   The estimated photon noise limits in the $J$, $H$, $K$ bands are 2.3$\times$10$^{-4}$, 
   2.4$\times$10$^{-4}$, and 3.7$\times$10$^{-4}$ per two-minute interval, respectively. This 
   uncertainties rescaling has barely changed the best-fit values, but enlarged their uncertainties, 
   thus decreased the detection significance. The derived jump parameters and coefficients for each 
   band are listed in Table~\ref{tab:result}, while the posterior joint probability distributions 
   between $T_{\rm{mid}}$ and $F_p/F_*$ are shown in Fig.~\ref{fig:corr}.
   
   In the second scenario, we changed the form of occultation model to $E(T_{58},T_{\rm{mid}},F_p/F_*)$ 
   so that we could fit the occultation duration $T_{58}$. Since our light curves are of poor quality, 
   we decided to adopt the PDFs of $T_{\rm{mid}}$ and $F_p/F_*$ from the light-curve modeling 
   in the first scenario as Gaussian priors input to the light-curve modeling in the second scenario. 
   Another five chains of 1\,000\,000 links were run in search for the best-fit occultation duration 
   of the $J$ and $K$ light curves. The $H$ band was not fitted because there was no detection. We 
   obtained an occultation duration of $0.1001^{+0.0063}_{-0.0070}$\,days for the $J$ band and 
   $0.1026^{+0.0054}_{-0.0058}$\,days for the $K$ band. They are both consistent with the primary 
   transit duration $T_{14}$=0.1004\,days \citep[derived using parameters of][]{2011PASJ...63..287F} 
   within their large uncertainties.
   
   In the third scenario, we tried to model the light curves from the nodding observation (the Jul-26-2011 
   dataset) to directly compare the staring and nodding mode observations. The $J$ and $H$ bands in the 
   nodding mode could fail to be fitted because of their extremely poor data quality. Thus we only modeled 
   the $K$ band. The light curve was divided into four groups of nods according to their nodding positions. 
   In the modeling, all four sub-light-curves share the same $F_p/F_*$ and have the mid-eclipse time fixed 
   on the expected mid-point assuming zero eccentricity, while they are allowed to have different coefficients 
   in the baseline from nod to nod. The adopted baseline function is
   \begin{equation}
     B_K=c_0+c_1x+c_2y+c_3xy+c_4x^2+c_5y^2.
   \end{equation}
   We performed the MCMC-based modeling in the same manner as in the first scenario to find the scaling factors 
   ($\beta_1$=1.47 and $\beta_2$=1.49) and to determine the flux ratio. This resulted in a flux ratio 
   of 0.268$\pm$0.076\% and $0.27^{+0.16}_{-0.15}$\% for the unscaled and rescaled versions, respectively. 
   The $rms$ of O--C residuals for this nodding light-curve is 4214~ppm per two-minute interval, twice 
   as high as the staring mode. Considering that the nodding observations featured crashes, a better 
   comparison would be using the un-crashed nodding pair. The $rms$ for the first 2-hour parts of nod A 
   and B is 4738~ppm, while for the last 1-hour parts of nod C and D it is 2638~ppm. The main difference 
   between these two un-crashed nodding pairs is their nodding pattern, that is, the locations of the 
   same star on the detector are different, which results in instrumental systematics of very different 
   levels. A location change also exists within one nodding pair. In contrast, the star's location on the 
   detector is relatively stable in the staring observation. Without the risk of introducing unexpected 
   systematics from a different location, it is easier to model the light curve, which results in 
   higher precision. This reconfirms that the staring mode is a better suited strategy than the nodding mode 
   in exoplanet observations, which has been noted in several previous observations 
   \citep[e.g. for \object{TrES-3b}:][]{2009A&A...493L..35D,2010ApJ...718..920C}.
   
   In addition to this analysis, we also examined our light curves to determine the correlations between 
   measured eclipse depth and the choices of aperture size and reference ensemble (see Fig.~\ref{fig:app_fig1} 
   and \ref{fig:app_fig2} in the appendix). As the aperture radius increases, the $rms$ of light-curve O--C 
   residuals first decreases to a minimum and then rises, which is expected because smaller apertures might 
   lose partial stellar flux while larger apertures would include more sky noise. Correspondingly, the 
   measured eclipse depth first changes greatly with the aperture size and then stabilizes when the aperture 
   size approaches our chosen value. For aperture sizes that result in $rms$ similar to that of the chosen 
   aperture, the measured eclipse depths agree well with our reported result within 1-$\sigma$ error bars. 
   Furthermore, the measured eclipse depths derived from different combinations of reference stars are 
   consistent with each other when they produce light curves with relatively low red noise. Therefore, we 
   confirm that our choices of photometry and reference ensemble are ideal, in contrast, the measured 
   eclipse depths are relatively insensitive to the choice of aperture size and reference ensemble. 
   
   \begin{figure}
      \centering
      \includegraphics[width=\hsize]{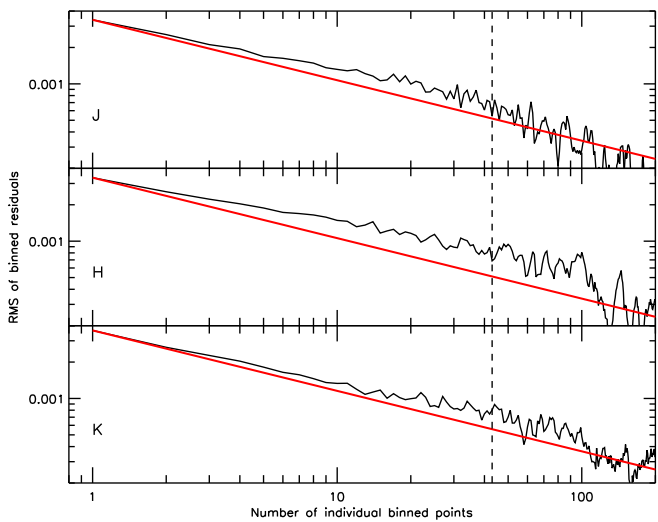}
      \caption{$RMS$ of the binned residuals v.s. bin size for the $J$, 
            $H$ and $K$ occultation light-curves. The red lines shows
            the prediction for Gaussian white noise ($1/\sqrt{N}$). 
            The vertical dashed line show the corresponding ingress/egress 
            duration time. We can still see the strong effect of correlated 
            noise in all three bands even after the baseline correction.\label{fig:rms}}
   \end{figure}
  
   \begin{table}
     \caption{Adopted parameters in the modeling process\label{tab:adopt}}
     \centering
     \begin{tabular}{lccc}
     \hline\hline
       Parameter & Units & Value & Ref.\\
     \hline
         T$_{\rm{mid,tran}}$ & BJD$_{\rm{TDB}}$ & 2454375.62510 & a\\
         Period & days   & 1.62843142 & a\\
    Inclination & degree & 85.58      & a\\
        $a/R_*$ & ...    & 5.37       & a\\
 $R_{\rm{p}}/R_*$ & ...  & 0.1108     & a\\
   Eccentricity & ...    & 0.         & a\\
       $\omega$ & ...    & 0.         & a\\
     $\log g_*$ & cgs    & 4.395 $^{+0.043}_{-0.040}$ & b\\
 $T_{\rm{eff}}$ & K      & 5700$\pm$100   & b\\
     $[$Fe/H$]$ & dex    & +0.09$\pm$0.09 & b\\
     \hline
     \end{tabular}
     \tablefoot{
                \tablefoottext{a}{\citet{2011PASJ...63..287F}}
                \tablefoottext{b}{\citet{2009A&A...496..259G}}
               }
   \end{table}
   
\section{Results and discussions}\label{sec:discuss}

   We list the derived jump parameters of our MCMC analysis in Table~\ref{tab:result}, 
   along with the coefficients of the baseline functions from the modeling in the first 
   scenario. Figure~\ref{fig:nod} shows the $K$-band light curve from the nodding observation 
   (Jul-26-2011), while Fig.~\ref{fig:occ} shows all three light curves from the staring 
   observation (Sep-08-2011). The adopted parameters that were used in our MCMC analysis 
   are given in Table~\ref{tab:adopt}.

   \subsection{Correlated noise}
   In our MCMC analysis, we have propagated the uncertainties of the baseline detrending 
   function into the PDFs of jump parameters by solving its coefficients at each MCMC 
   step. We also tried to account for the red noise by rescaling the photometric 
   uncertainties with the $\beta$ factors. The time averaging processes are shown in 
   Fig.\ref{fig:rms}. The $rms$ of the binned residuals clearly deviates from the 
   predicted Gaussian white noise, as shown by the red lines, indicating the presence 
   of correlated noise in our light curves. Here we employed another commonly used 
   method \citep[e.g.:][]{2008MNRAS.386.1644S}, the "prayer-bead" residual permutation 
   method, which preserves the shape of time-correlated noise, to investigate whether 
   there is still excess red noise that is not included in our MCMC analysis. Firstly, 
   the best-fit model was removed from the light curve. The residuals were cyclically 
   shifted from the $i$th to the $i$+1th positions (of the $N$ data points), while 
   off-position data points in the end were wrapped at the beginning. Then the best-fit 
   model was added back to the permuted residuals to form a new synthetic light curve. 
   This synthetic light curve was then fitted in the same way as the real light curve, 
   as described in Sect.~\ref{sec:lc}. We inverted the light-curve sequence to perform 
   another series of cycling, thus achieving 2$N$$-$1 synthetic light curves in 
   total. We also calculated the median and 68.3\% confidence level of the resulting 
   distribution as the best-fit value and $1\sigma$ uncertainties. 
   
   This residual-permutation (RP)-based analysis leads to a flux ratio of 
   $0.268^{+0.062}_{-0.054}$\% for the $K$ band and $0.167^{+0.033}_{-0.038}$\% for 
   the $J$ band. The RP-based flux ratios have smaller uncertainties than those of 
   the $\beta$-based MCMC analysis (0.269$\pm$0.062\%  and $0.168^{+0.050}_{-0.052}$\%, 
   correspondingly). The differences in best-fit values are very small. This 
   indicates that our $\beta$-based MCMC analysis has included the potential impact 
   from the time-correlated noise. We adopted the $\beta$-based MCMC results as our 
   final results.
   
   \subsection{Orbital eccentricity}
   We obtain an average mid-occultation offset time of 10.5$\pm$3.1 minutes and an average 
   occultation duration time of $0.1016^{+0.0041}_{-0.0045}$ days by combining values 
   of the $J$ and $K$ bands with weights according to the inverse square of their 
   uncertainties. The secondary eclipse of \object{WASP-5b} is expected to occur at 
   phase $\phi$=0.5002 if it is in a circular orbit. This value has taken into account 
   the delayed light travel time of $\sim$27~s \citep{2005ApJ...623L..45L}. However, 
   our average mid-occultation time occurs at a delayed offset of 10.1$\pm$3.1 minutes 
   to this expected phase, which might indicate a nonzero eccentricity. We used equations 
   from \citet{2009ApJ...698.1778R} to derive the values of $e\cos\omega$ and $e\sin\omega$: 
   \begin{equation}
     e\cos\omega\simeq\pi\Delta\phi/2
   \end{equation}
   \begin{equation}
     e\sin\omega=\frac{D_{\mathrm{II}}-D_{\mathrm{I}}}{D_{\mathrm{II}}+D_{\mathrm{I}}}\frac{\alpha^2-\cos^2i}{\alpha^2-2\cos^2i},
   \end{equation}
   where $\alpha$=$(R_*/a+R_{\rm{p}}/a)/\sqrt{1-e^2}$, while $D_{\rm{II}}$ and $D_{\rm{I}}$ 
   refer to the durations of secondary eclipse and primary transit. Thus $e\cos\omega$ and 
   $e\sin\omega$ can be constrained if we can measure the mid-eclipse time and duration of 
   a secondary eclipse with sufficient precision. We calculated an $e\cos\omega$=0.0067$\pm$0.0021 
   and an $e\sin\omega$=0.007$\pm$0.026. While the former parameter barely deviates from zero 
   by $3\sigma$, the latter is consistent with zero within its large uncertainty. The 68.3\% 
   confidence level for eccentricity is $e$=0.020$^{+0.019}_{-0.011}$, with corresponding 
   argument of periastron $|\omega|$=71$^{+11\circ}_{-31}$. Our derived eccentricity is only 
   slightly larger than zero at a significance lower than $2\sigma$. From the previous radial 
   velocity studies, \citet{2009A&A...496..259G} found a tentatively nonzero value of 
   $e$=0.038$^{+0.026}_{-0.018}$, while \citet{2012MNRAS.422.3151H} claimed that its eccentricity 
   is compatible with zero ($e$=0.012$\pm$0.007) based on more RV measurements. Our result, 
   derived from a different approach, lies between them and is consistent with both results 
   within their errorbars. Recent Warm {\it Spitzer} measurements resulted in a mean value of 
   $e\cos\omega$=0.0025$\pm$0.0012 \citep{2013ApJ...773..124B}, which is 1.74$\sigma$ lower than 
   our average value (c.f. 0.86$\sigma$ lower than our $K$-band result, see Table~\ref{tab:result}).
   However, we are cautious to draw any conclusion on nonzero eccentricity here. The shapes 
   of our light curves are complicated due to the existence of instrumental and atmospheric 
   systematics, as can be seen in Fig.\ref{fig:occ}. It is very likely that these systematic 
   effects bias the mid-eclipse time. Furthermore, the occultation duration is poorly constrained 
   by our measurements.

   \subsection{Eclipse depths and brightness temperatures}\label{sec:brightness_temp}      
   
  To preliminarily probe the atmosphere, we first calculated the brightness temperatures 
  corresponding to the measured flux ratios. We assumed blackbody emission for the planet 
  and interpolated the stellar spectrum in the Kurucz stellar models \citep{1979ApJS...40....1K} 
  for the host star (using $T_{\rm{eff}}$=5700\,K, $\log g$=4.395 and [Fe/H]=0.0). The blackbody 
  spectrum and the stellar spectrum were both integrated over the bandpass of our three NIR 
  bands individually. The blackbody temperature that yields the resulting flux ratio best-fit 
  was adopted as the corresponding brightness temperature in each band. For the Sep-08-2011 
  secondary eclipse, the measured flux ratios are $0.168^{+0.050}_{-0.052}$\% and 0.269$\pm$0.062\% 
  in the $J$ and $K$ band, and a $3\sigma$ upper limit of 0.166\% in the $H$ band, which 
  translates to brightness temperatures of $2996^{+212}_{-261}$\,K, $2890^{+246}_{-269}$\,K 
  and $<2779$\,K ($3\sigma$), respectively.  
  
  We used the same approach to calculate the brightness temperature of the Warm 
  {\it Spitzer} data, where \citet{2013ApJ...773..124B} reported 0.197$\pm$0.028\% at 
  3.6~$\mu$m and 0.237$\pm$0.024\% at 4.5~$\mu$m. In this way, the temperatures in 
  the NIR and MIR were derived with the same stellar atmosphere models and parameters 
  \citep[c.f.][]{2013ApJ...773..124B}. As a result, the {\it Spitzer} eclipse depths 
  translate into brightness temperatures of 1982$^{+117}_{-122}$~K and 1900$^{+92}_{-94}$~K, 
  respectively. The temperature derived from our NIR data ($\sim$2700~K) completely 
  disagrees with that derived from the Warm {\it Spitzer} data ($\sim$1900~K). 
  
  \subsection{Constraints on atmospheric properties}\label{sec:atmos}
 
  \subsubsection{Atmospheric models}
  To investigate possible scenarios of the atmospheric properties, we modeled the emerging 
  spectrum of the dayside atmosphere of \object{WASP-5b} using the exoplanetary atmospheric 
  modeling and retrieval method of \citet{2009ApJ...707...24M,2010ApJ...725..261M}. Our model 
  performs line-by-line radiative transfer in a plane-parallel atmosphere, with constraints 
  on local thermodynamic equilibrium, hydrostatic equilibrium, and global energy balance. The 
  pressure-temperature ($P$-$T$) profile and the molecular composition are free parameters of 
  the model, allowing exploration of models with and without thermal inversions, and with 
  oxygen-rich as well as carbon-rich compositions \citep{2012ApJ...758...36M}. The model 
  includes all the primary sources of opacity expected in hydrogen-dominated atmospheres in 
  the temperature regimes of hot Jupiters, such as molecular line absorption due to various 
  molecules (H$_2$O, CO, CH$_4$, CO$_2$, HCN, C$_2$H$_2$, TiO, VO) and collision-induced 
  absorption (CIA) due to H$_2$ \citep[see][for more details]{2012ApJ...758...36M}. The 
  volume-mixing ratios of all the molecules are free parameters in the model. Given that 
  the number of model parameters ($N$=10-14, depending on the C/O ratio) is much higher than 
  the number of available data points, our goal is to nominally constrain the regions of model 
  space favored by the data rather than determine a unique fit.  

  \subsubsection{Constraints from our NIR data}
\begin{figure*}
     \centering
     \includegraphics[width=0.48\textwidth]{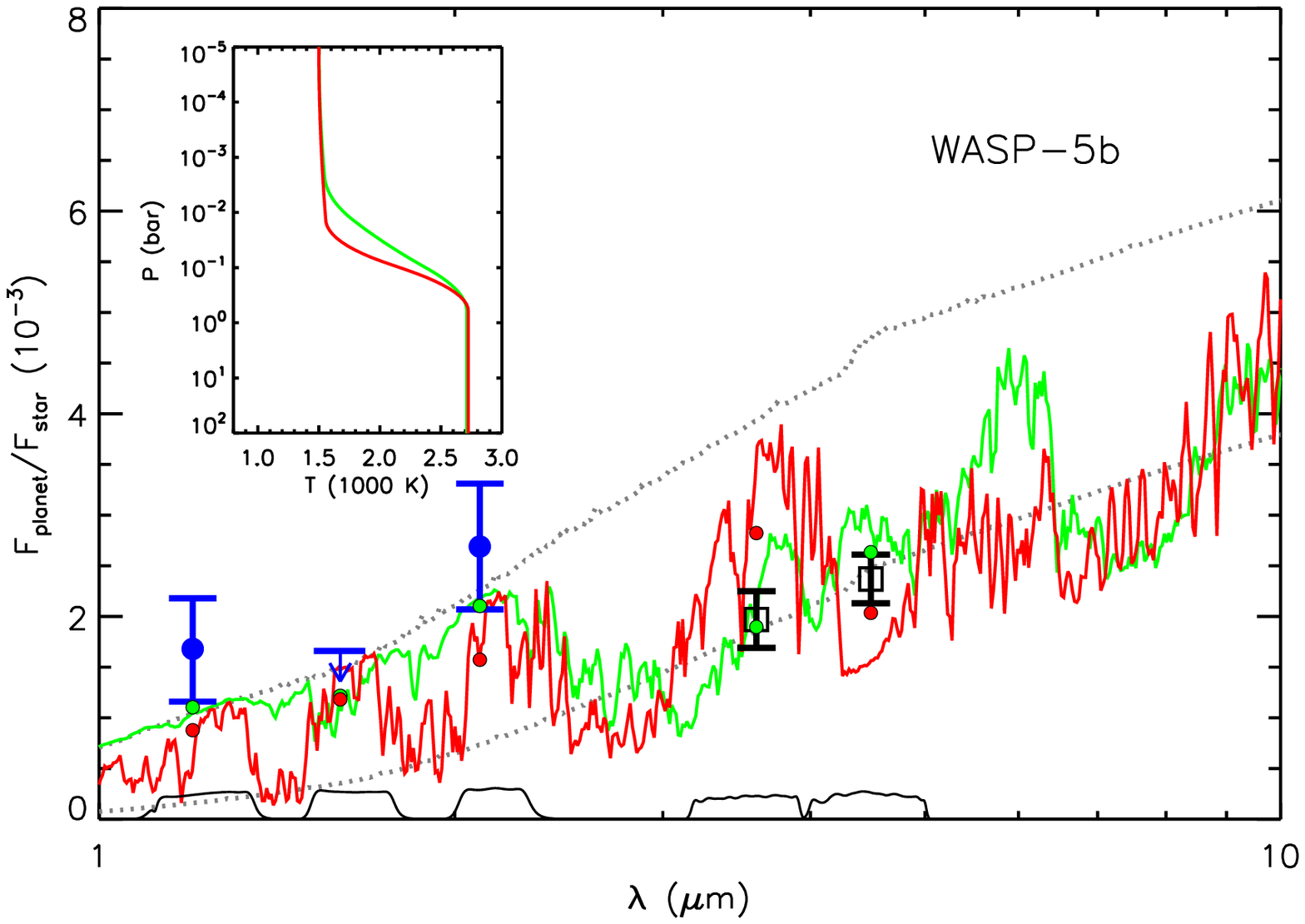}
     \includegraphics[width=0.48\textwidth]{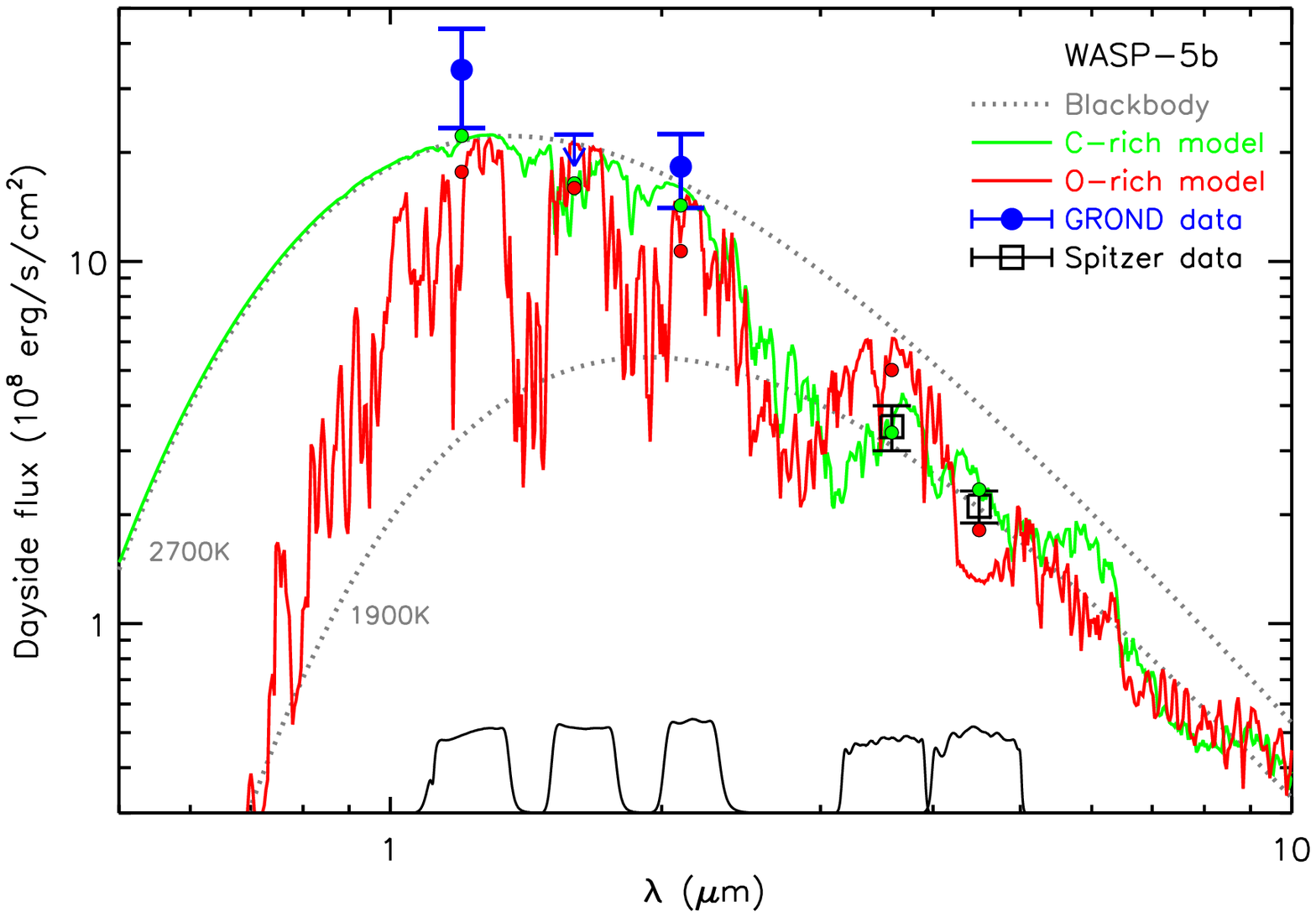}
     \caption{Dayside thermal emission spectrum from the hot Jupiter \object{WASP-5b} in terms of 
              planet-to-star flux ratios ({\it left}) and planetary dayside flux ({\it right}). The 
              blue circles with error bars at 1.26~$\mu$m and 2.15~$\mu$m and the upper-limit at 
              1.65~$\mu$m,  show our measured planet-star flux ratios in the photometric $J$, $H$, 
              and $K$ bands. The {\it Spitzer} photometric observations reported by \citet{2013ApJ...773..124B} 
              are shown at 3.6~$\mu$m and 4.5~$\mu$m (black squares). The dotted gray curves show 
              two planetary blackbody spectra with temperatures of 1900~K and 2700~K. While our 
              $J$, $H$, $K$ band data can be fit with a $\sim$2700~K blackbody spectrum, the 
              {\it Spitzer} data is consistent with a $\sim$1900~K blackbody. The red and green 
              curves show two model spectra of the dayside atmosphere of \object{WASP-5b} with 
              oxygen-rich and carbon-rich compositions, respectively. The inset shows the corresponding 
              pressure-temperature profiles without thermal inversions; a thermal inversion in the 
              dayside atmosphere of \object{WASP-5b} is ruled out by the data. An oxygen-rich model 
              is unable to explain all the data. On the other hand, while a carbon-rich model provides 
              a good fit to the data, it radiates 70\% more energy than the incident irradiation, 
              which may be unphysical unless there are additional absorbing species in the 
              atmosphere that are not accounted for in the current C-rich model. See 
              Sect.~\ref{sec:atmos} for discussion.}\label{fig:emm} 
   \end{figure*}

  Our observations place a stringent constraint on the temperature structure of the lower atmosphere 
  of the planetary dayside. The $J$, $H$, and $K$ bands contain only weak molecular features due to 
  spectroscopically dominant molecules in hot-Jupiter atmospheres. As such, photometric observations 
  in these bands probe deep into the lower regions of the planetary atmosphere until the high pressures 
  make the atmosphere optically thick (around $P \sim 0.1-1$~bar) due to H$_2$-H$_2$ CIA continuum 
  absorption \citep{2012ApJ...758...36M}. Our observed brightness temperatures in the $J$, $H$, and 
  $K$ bands can be explained by a roughly isothermal temperature profile of $\sim$2700~K in the 
  lower atmosphere of \object{WASP-5b}, consistent with the fact that for highly irradiated hot Jupiters 
  the dayside temperature structure at $\tau$$\sim$1 tends to be isothermal \citep{2008ApJS..179..484H,
  2009ApJ...707...24M,2010A&A...520A..27G}. In principle, our $J$ and $K$ band data allow for 
  significantly higher temperatures, up to $\sim$3200~K, but our $H$-band observation rules out 
  temperatures above $\sim$2700~K. As shown in Fig.~\ref{fig:emm}, a blackbody spectrum of 2700~K 
  representing the continuum blackbody of the lower atmosphere provides a reasonable fit to the $J$, 
  $H$, $K$ data.

  However, an isothermal temperature profile at $\sim$2700~K over the entire vertical extent of 
  the atmosphere is unlikely. This would violate global energy balance since the planet would radiate 
  substantially more energy than it receives. We assume that the internal source of energy is negligible 
  compared to the incident irradiation \citep{2008ApJ...678.1436B}. 

  \subsubsection{Constraints from NIR and {\it Spitzer} data}
  Additional constraints on the temperature profile and on the chemical composition of the dayside 
  atmosphere of \object{WASP-5b} are obtained by combining our data with new photometric observations 
  obtained with the {\it Spitzer Space Telescope} at 3.6~$\mu$m and 4.5~$\mu$m \citep{2013ApJ...773..124B}. 
  
  Our data together with the {\it Spitzer} data rule out a thermal inversion in \object{WASP-5b} 
  irrespective of its chemical composition, consistent with the finding of \citet{2013ApJ...773..124B} 
  based on the {\it Spitzer} data alone. The {\it Spitzer} data probe higher atmospheric layers 
  than the ground-based data due to strong molecular absorption in the two {\it Spitzer} bands, 
  and are consistent with brightness temperatures of $\sim$1900~K, which is much lower than the 
  $\sim$2700~K temperatures in the ground-based channels. Consequently, the two data sets suggest 
  temperatures decreasing outward in the atmosphere. 
  
  Previous work has shown that the {\it Spitzer} data also provide good diagnostics of the C/O ratio 
  of the atmosphere as the bandpasses overlap with broad spectroscopic features of several dominant 
  C- and O-bearing molecules \citep{2012ApJ...758...36M}. Chemical compositions of hot-Jupiter 
  atmospheres can be extremely different depending on whether they are oxygen-rich (C/O $<$ 1) or 
  carbon-rich (C/O$\geq$1). Whereas in O-rich atmospheres (e.g. of solar composition, with C/O = 0.5), 
  H$_2$O and CO, and possibly TiO and VO, are the dominant sources of opacity, C-rich atmospheres are 
  depleted in H$_2$O and abundant in CO, CH$_4$, HCN, and C$_2$H$_2$ \citep{2011Natur.469...64M,
  2012ApJ...745...77K,2012ApJ...758...36M,2013ApJ...763...25M}. 

  We investigated both O-rich and C-rich scenarios in the present work and found that neither composition 
  simultaneously provides a good fit to the data and satisfies energy balance, as shown in Fig.~\ref{fig:emm}. 
  However, the chemical composition is poorly constrained by the current data. First, we found that an O-rich 
  solar composition atmosphere can neither fit all the data to within the 1-$\sigma$ errors nor satisfy 
  energy balance; it radiates more energy than it receives. A composition with enhanced metallicity 
  ($5\times$solar), but still O-rich, can satisfy energy balance, but still does not fit all the data, 
  predicting the planet-star flux contrast at 3.6~$\mu$m to be $\gtrsim$3-$\sigma$ higher than the observed 
  value, as shown in Fig.~\ref{fig:emm}. On the other hand, a C-rich model can fit all the data reasonably
  well, but radiates 70\% more energy than it receives from incident radiation, thereby violating global energy 
  balance. 
  
  Both the O-rich and C-rich models were consistent with the lack of a thermal inversion in the planet.
  While the high chromospheric activity of the host star could destroy inversion-causing species in the 
  atmosphere irrespective of its C/O ratio \citep{2010ApJ...720.1569K,2013ApJ...773..124B}, a C-rich 
  atmosphere would be naturally depleted in oxygen-rich inversion-causing compounds such as TiO 
  and VO, which designates \object{WASP-5b} as a C2-class hot Jupiter in the classification of 
  \citet{2012ApJ...758...36M}. 

  \subsubsection{Possible scenarios and future prospects}
  Our data agree with two possible scenarios for \object{WASP-5b}: a carbon-rich and an oxygen-rich 
  atmosphere. However, we caution that new observations are required to conclusively constrain its chemical 
  composition.
  
  The C-rich scenario, while providing a good fit to all available data, requires an explanation for 
  the apparent energy excess in the emergent spectrum. This could be mitigated by an additional absorber 
  in the atmosphere, which has high opacity blueward of the $J$ band ($\lesssim$1.1~$\mu$m) and/or a 
  strong feature in the $H$ band ($\sim$1.5--1.8~$\mu$m). The presence of such a component is currently 
  merely speculative, but could be seen or ruled out using follow-up spectroscopic observations, for 
  instance with {\it HST}/WFC3. Such observations would additionally constrain the energy budget of 
  the dayside atmosphere of \object{WASP-5b}.
  
  The O-rich model satisfies global energy balance but does less well at simultaneously fitting the 
  $J$, $H$, $K$ data and the {\it Spitzer} 3.6~$\mu$m point. One explanation is that the planet shows 
  substantial temporal variability in its emerging spectrum, but the magnitude of the inferred variability 
  seems implausibly high. Another possibility is that different systematic effects between the ground-based 
  and {\it Spitzer} data bias the derived thermal emission measurements.
  
  Spectroscopy with {\it HST}/WFC3 in the 1.1--1.7~$\mu$m bandpass would allow us to conclusively constrain 
  the chemical composition of the atmosphere, since our two model spectra in Fig.~\ref{fig:emm} predict 
  very different spectral shapes in that bandpass. In addition, observations of thermal phase curves with 
  warm {\it Spitzer} \citep[e.g.][]{2009ApJ...703..769K} will also allow us to place stringent constraints 
  on the day-to-night energy redistribution, since all models fitting our current data predict extremely low 
  redistributions implying strong day-to-night thermal contrasts.

\section{Conclusions}\label{sec:con}

   We observed two secondary eclipses of \object{WASP-5b} simultaneously in the $J$, $H$ and 
   $K$ bands with GROND on the MPG/ESO 2.2 meter telescope, one in nodding mode and the other 
   in staring mode. Although we failed to extract useful results from the nodding-mode 
   observation due to the associated complicated systematics, we did measure the occultation 
   dips from the staring-mode observation with reasonable precision, reconfirming that 
   the staring mode is more suited than the nodding mode for exoplanet observations.
   
   We have successfully detected the thermal emission from the dayside of \object{WASP-5b} in 
   the $J$ and $K$ bands, with flux ratios of $0.168^{+0.050}_{-0.052}$\% and 0.269$\pm$0.062\%, 
   respectively. In the $H$ band we derived a 3-$\sigma$ upper limit of 0.166\%. The 
   brightness temperatures inferred from the $J$ and $K$ bands are consistent with each other
   ($2996^{+212}_{-261}$\,K and $2890^{+246}_{-269}$\,K, respectively), but the upper limit 
   in the $H$ band rules out temperatures above 2779\,K at $3\sigma$ level. While a slight 
   difference might exist, together they indicate a roughly isothermal lower atmosphere of 
   $\sim$2700~K. We modeled the GROND data together with the Warm {\it Spitzer} data using 
   the spectral retrieval technique, ruling out a thermal inversion. We fit our data with two 
   different models: an oxygen-rich atmosphere and a carbon-rich atmosphere. The O-rich model 
   requires a very low day-to-night-side heat redistribution but satisfies energy balance. 
   The C-rich model fits our data better, but violates energy balance in that it radiates 
   70\% more energy than it receives. To constrain the chemical composition of \object{WASP-5b} 
   and to distinguish atmospheric models, more observations in the NIR, in particular 
   spectroscopy, are required.

\begin{acknowledgements}
      We thank the referee Bryce Croll for his careful reading and helpful 
      comments that improved the manuscript. We acknowledge Timo Anguita for 
      technical support of the observations. G.C. acknowledges the Chinese Academy 
      of Sciences and the Max Planck Society for the support of doctoral training 
      in the program. N.M. acknowledges support from the Yale Center for Astronomy 
      and Astrophysics (YCAA) at Yale University through the YCAA prize 
      postdoctoral fellowship. H.W. acknowledges the support by NSFC grants 
      11173060, 11127903, and 11233007. This work is supported by the 
      Strategic Priority Research Program "The Emergence of Cosmological 
      Structures" of the Chinese Academy of Sciences, Grant No. XDB09000000.
      Part of the funding for GROND (both hardware and personnel) was 
      generously granted from the Leibniz-Prize to G. Hasinger (DFG grant 
      HA 1850/28-1).
\end{acknowledgements}


  \Online

  \begin{appendix}
  \section{Additional figures}
  In this appendix, we present figures that show the dependence of measured 
  eclipse depth on the choice of aperture radii and on the choice of different 
  reference star combinations. We also display the correlations between raw 
  light-curve flux and detrending parameters.
   
\begin{figure*}
\begin{center}
\includegraphics[width=0.4\textwidth]{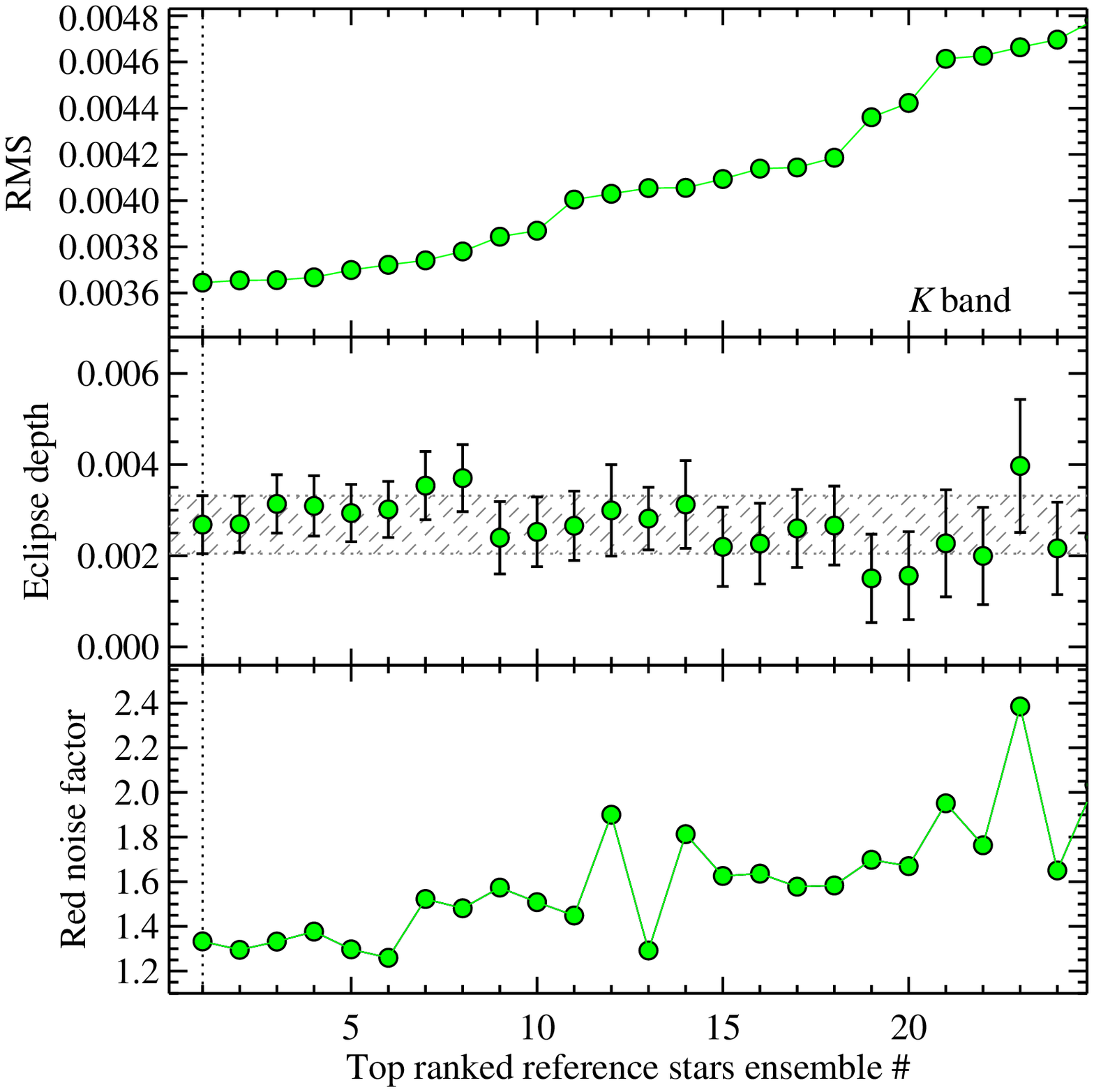}
\includegraphics[width=0.4\textwidth]{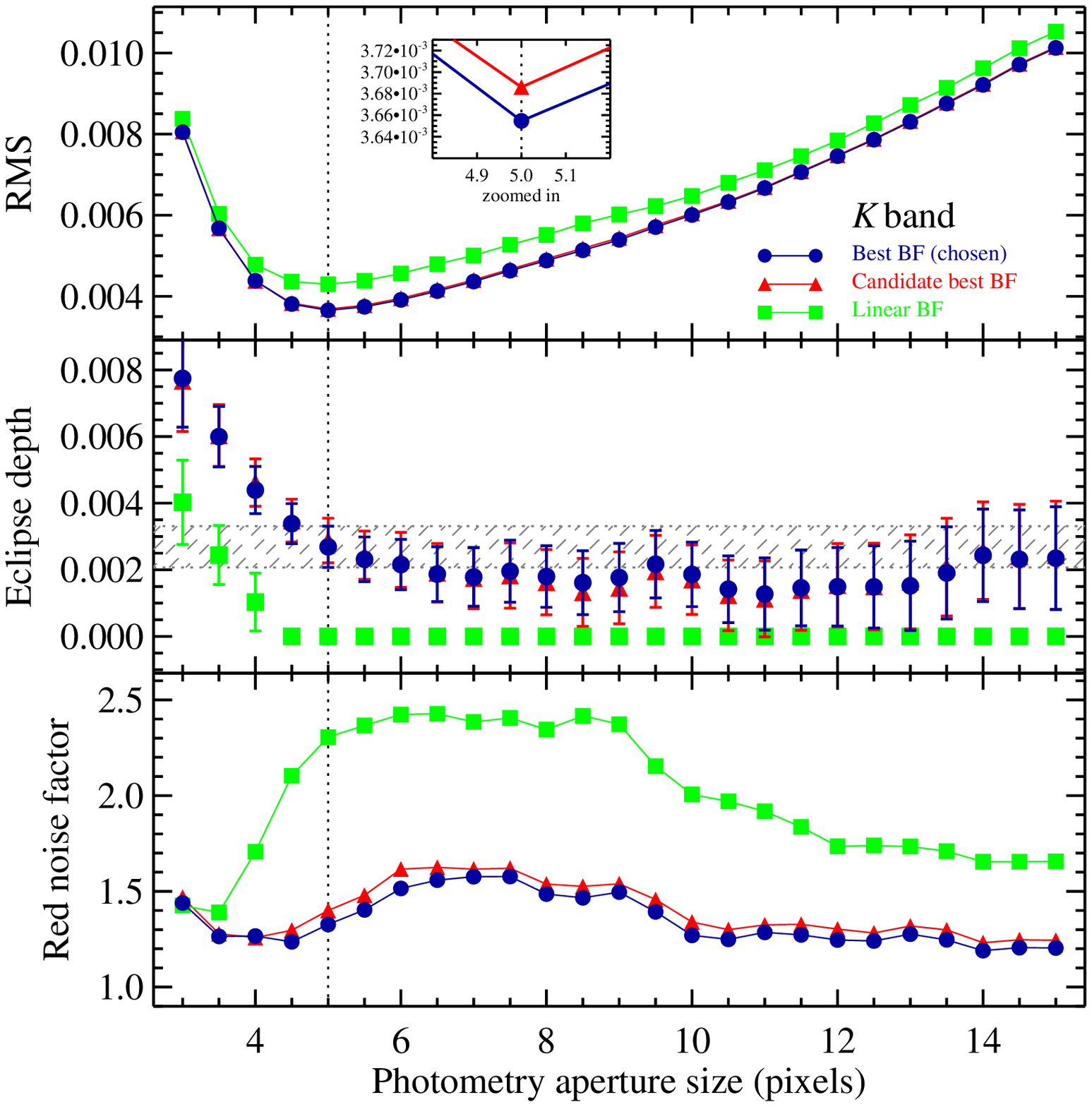}
\caption{Dependence of $rms$, measured eclipse depth, and red noise factor on aperture size and reference ensemble for the $K$ band. The {\it left panel} shows the dependence on aperture size. The top subpanel displays the $rms$ of light-curve O--C residuals. The middle subpanel displays corresponding eclipse depths. The bottom subpanel displays the time-averaging red noise factor (i.e. $\beta_2$ as denoted in Sect.~\ref{sec:lc}). The results derived from three cases of baseline functions (BF) are shown for comparison: the chosen best BF, a candidate BF that results in similar BIC to the best BF, and a linear BF that produces a poor fit. The vertical dotted line refers to the chosen aperture size, while the dash-shaded area refers to the 1-$\sigma$ confidence level of our reported result. The {\it right panel} shows the dependence of $rms$, measured eclipse depth, and red noise factor on different reference star ensembles. These ensembles have been sorted according to $rms$ for display purposes, with the best one being \#1. The subpanels on the right are organized in the same manner as for the left.}
\label{fig:app_fig1}
\end{center}
\end{figure*}

\begin{figure*}
\begin{center}
\includegraphics[width=0.4\textwidth]{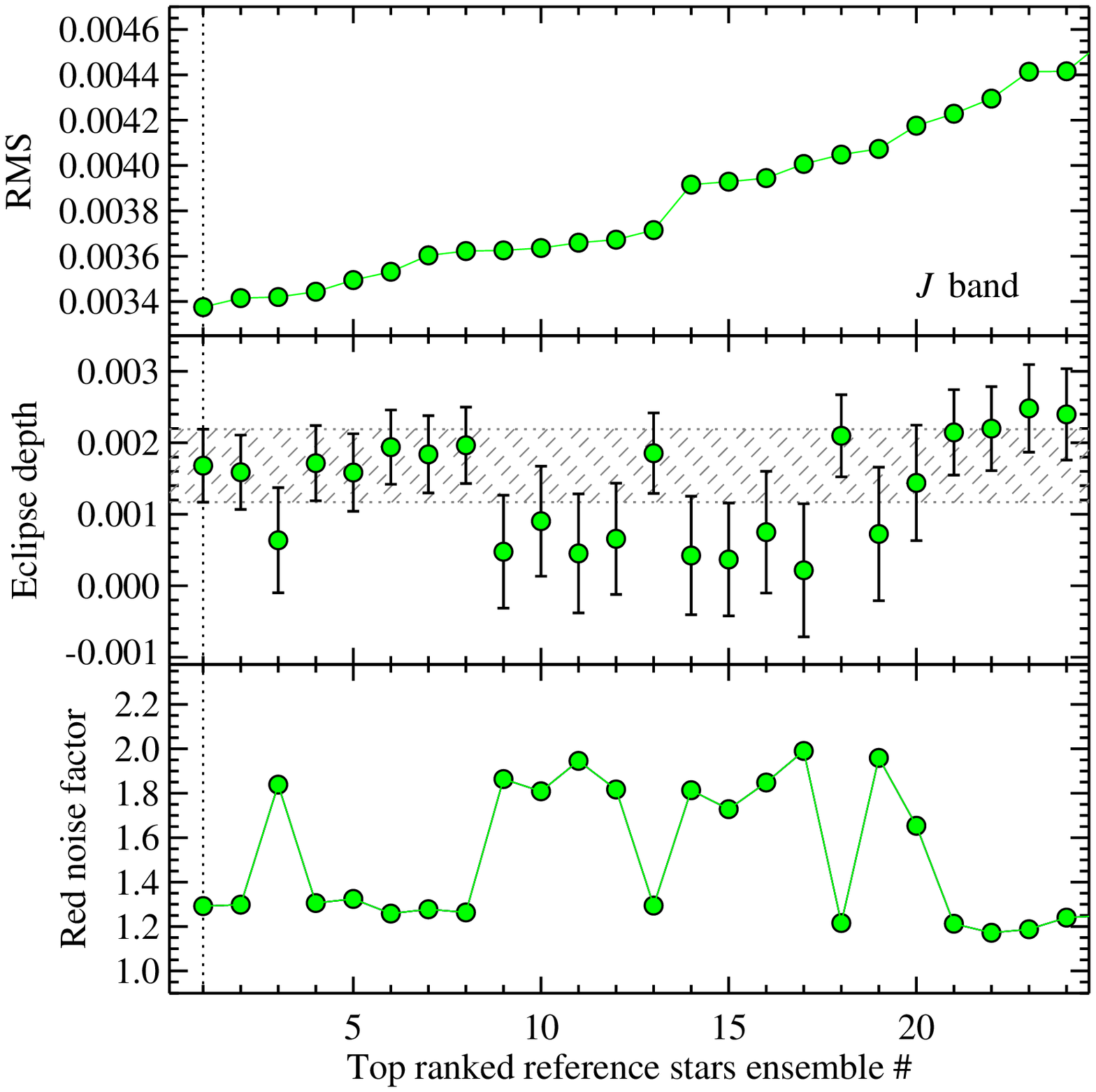}
\includegraphics[width=0.4\textwidth]{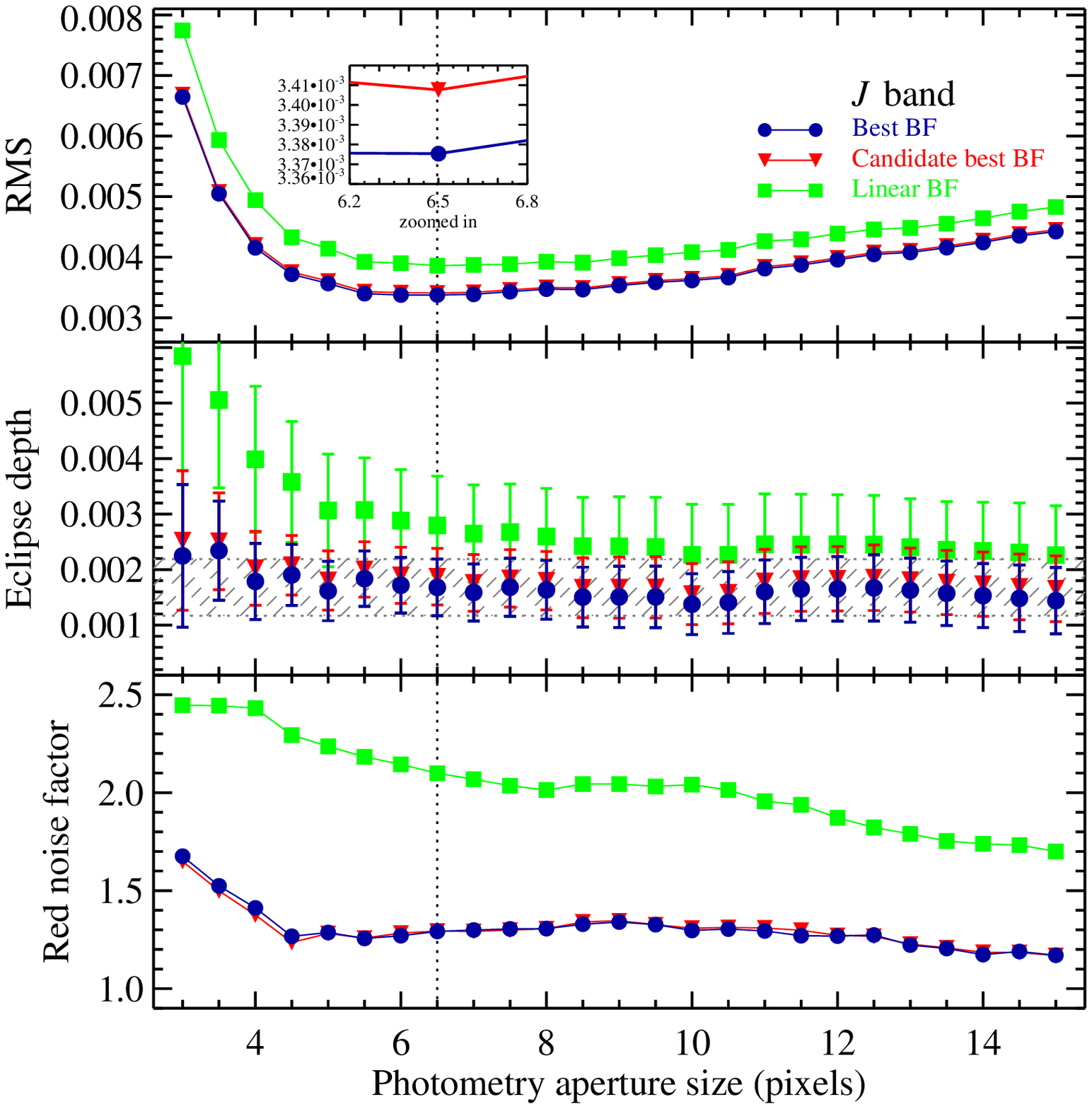}
\caption{Dependence of $rms$, measured eclipse depth, and red noise factor on aperture size and reference ensemble for the $J$ band. The subpanels in this figure are organized in the same manner as Fig.~\ref{fig:app_fig1}.}
\label{fig:app_fig2}
\end{center}
\end{figure*}

\begin{figure*}
\begin{center}
\includegraphics[width=0.24\textwidth]{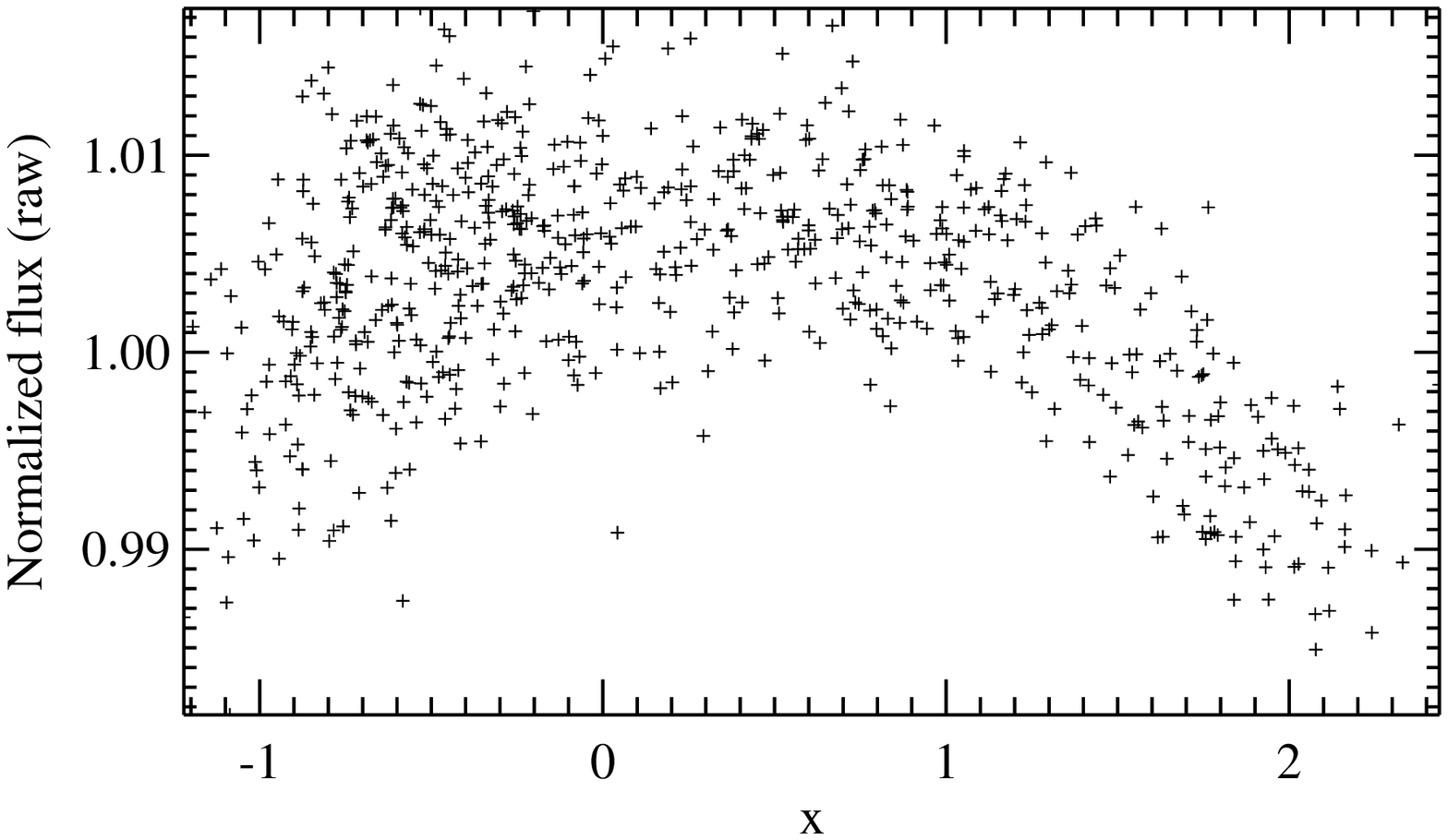}
\includegraphics[width=0.24\textwidth]{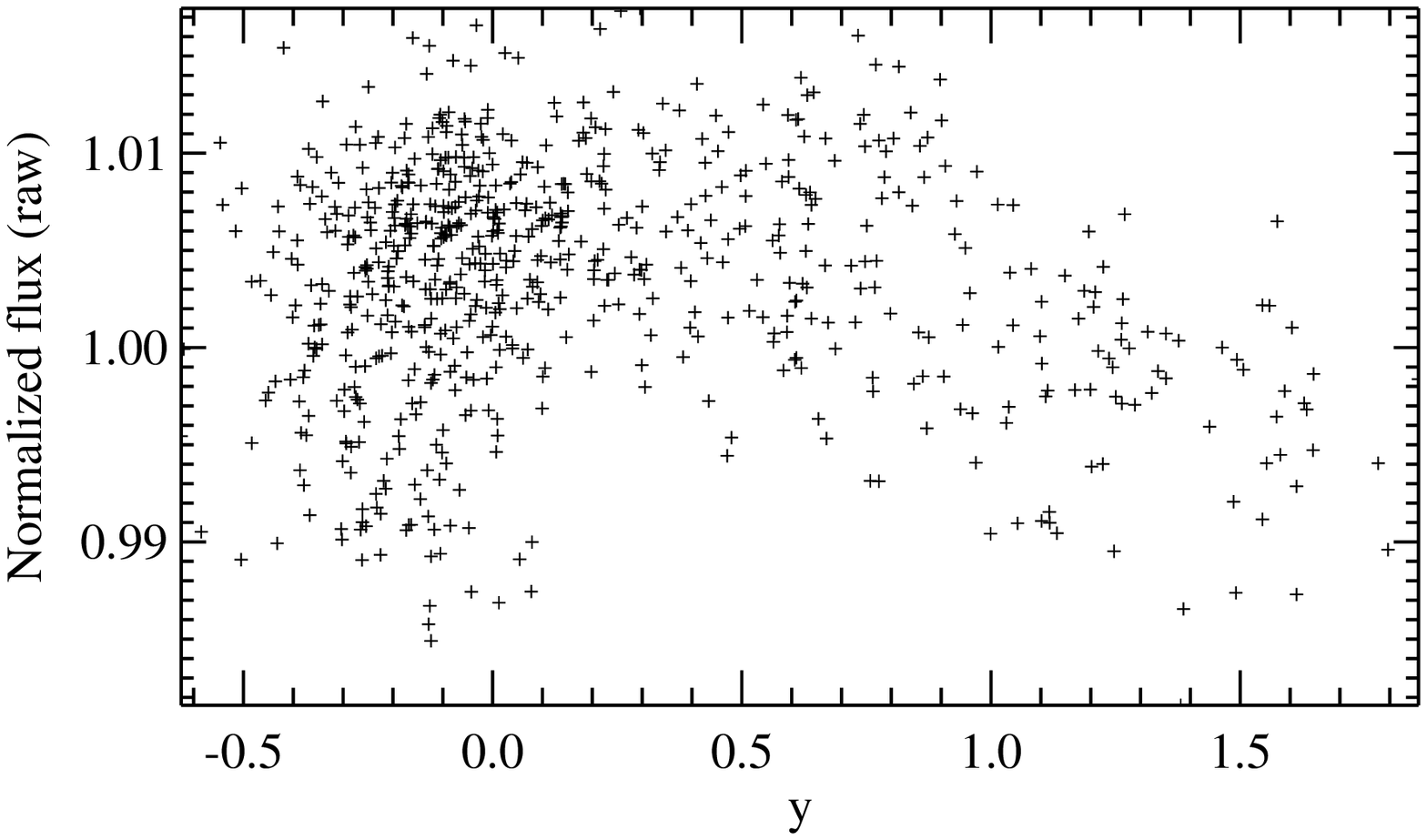}
\includegraphics[width=0.24\textwidth]{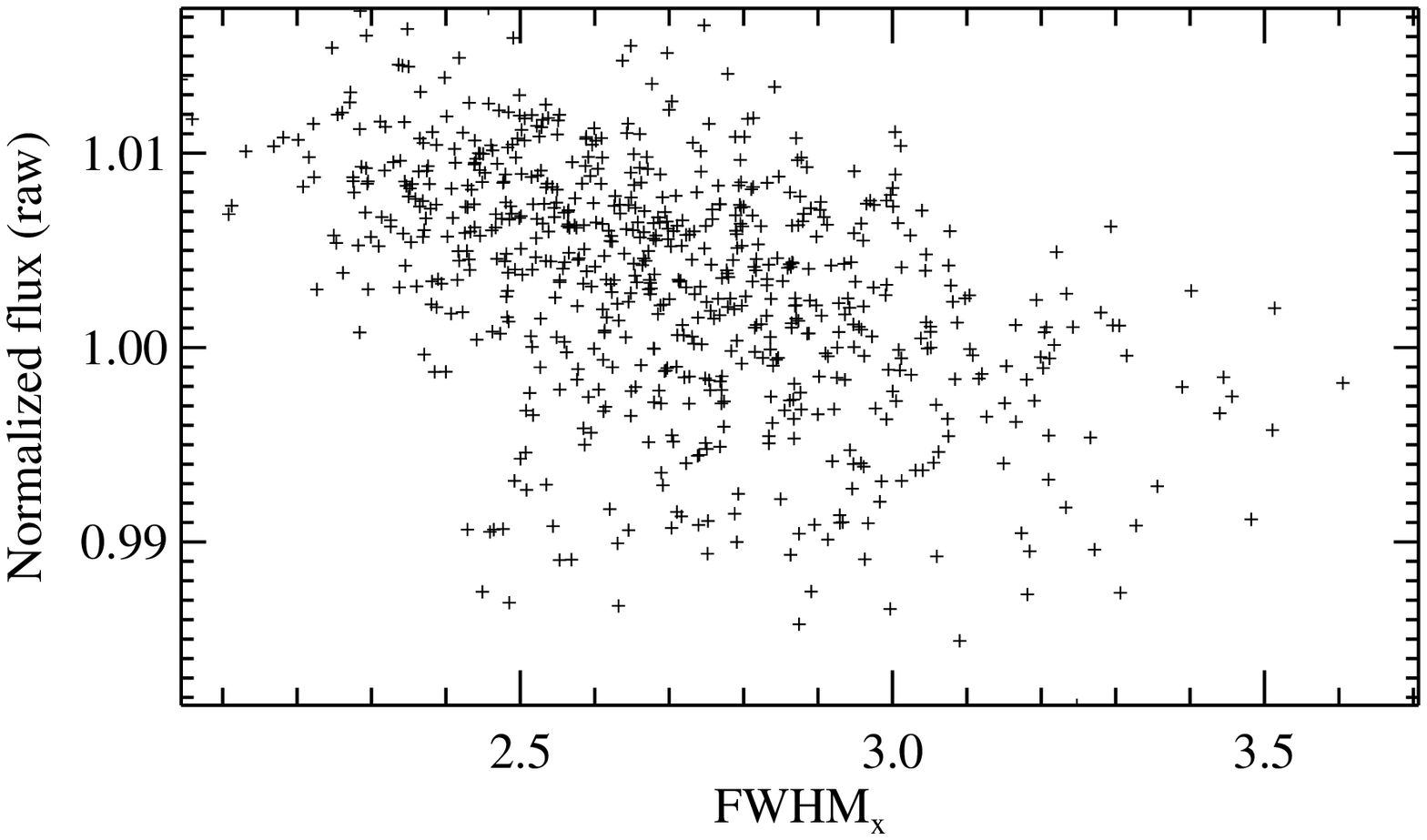}
\includegraphics[width=0.24\textwidth]{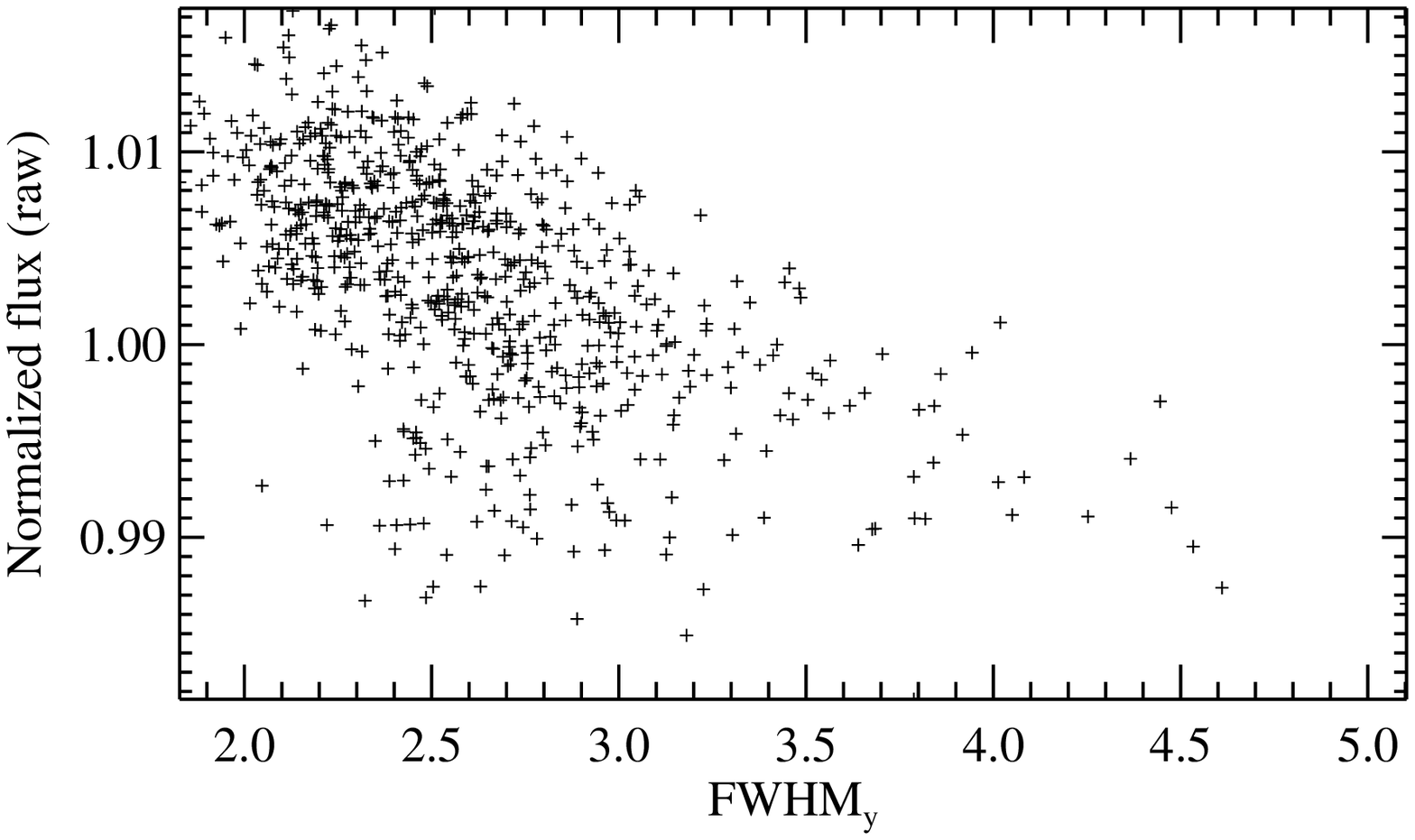}\\
\includegraphics[width=0.24\textwidth]{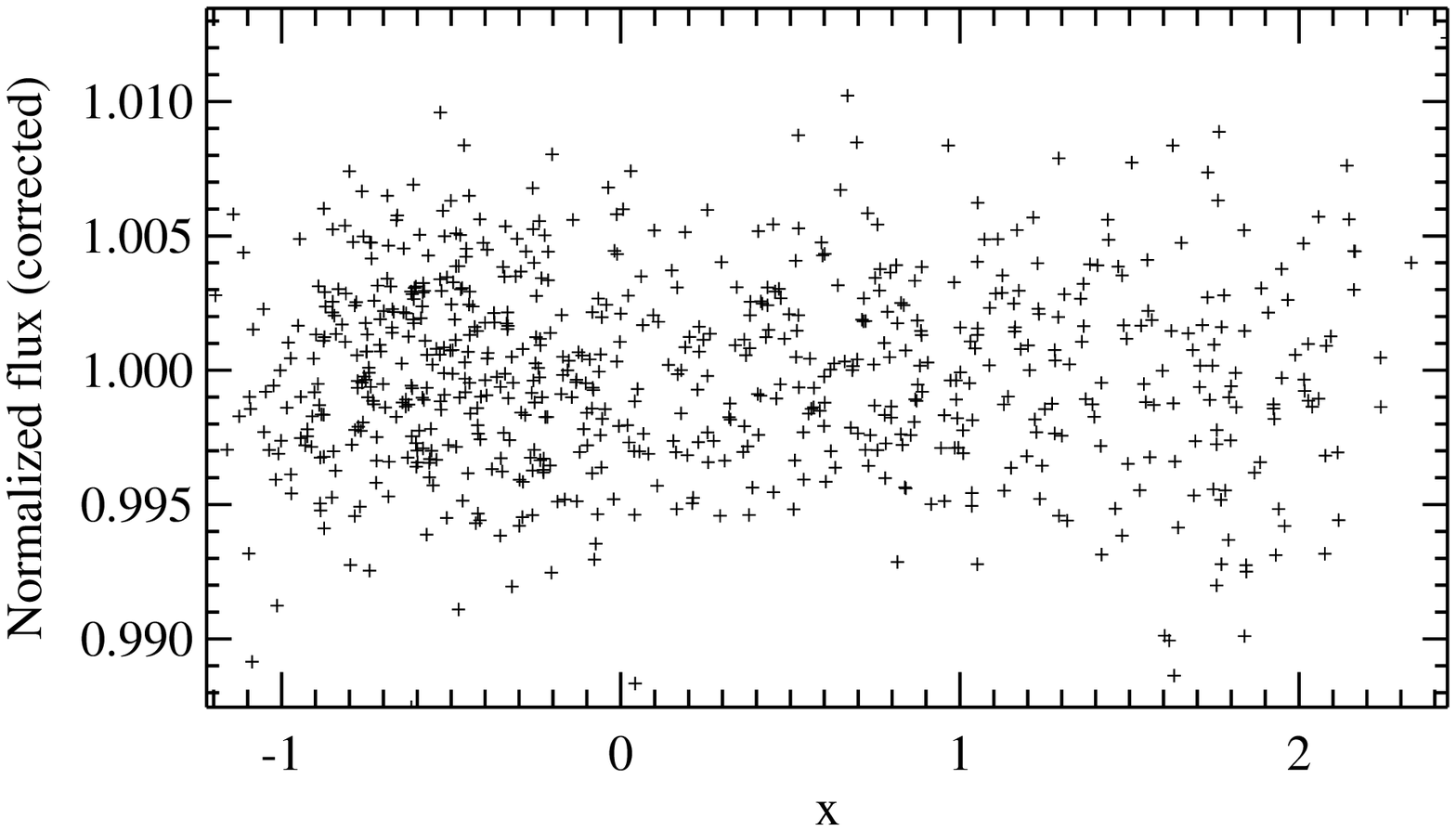}
\includegraphics[width=0.24\textwidth]{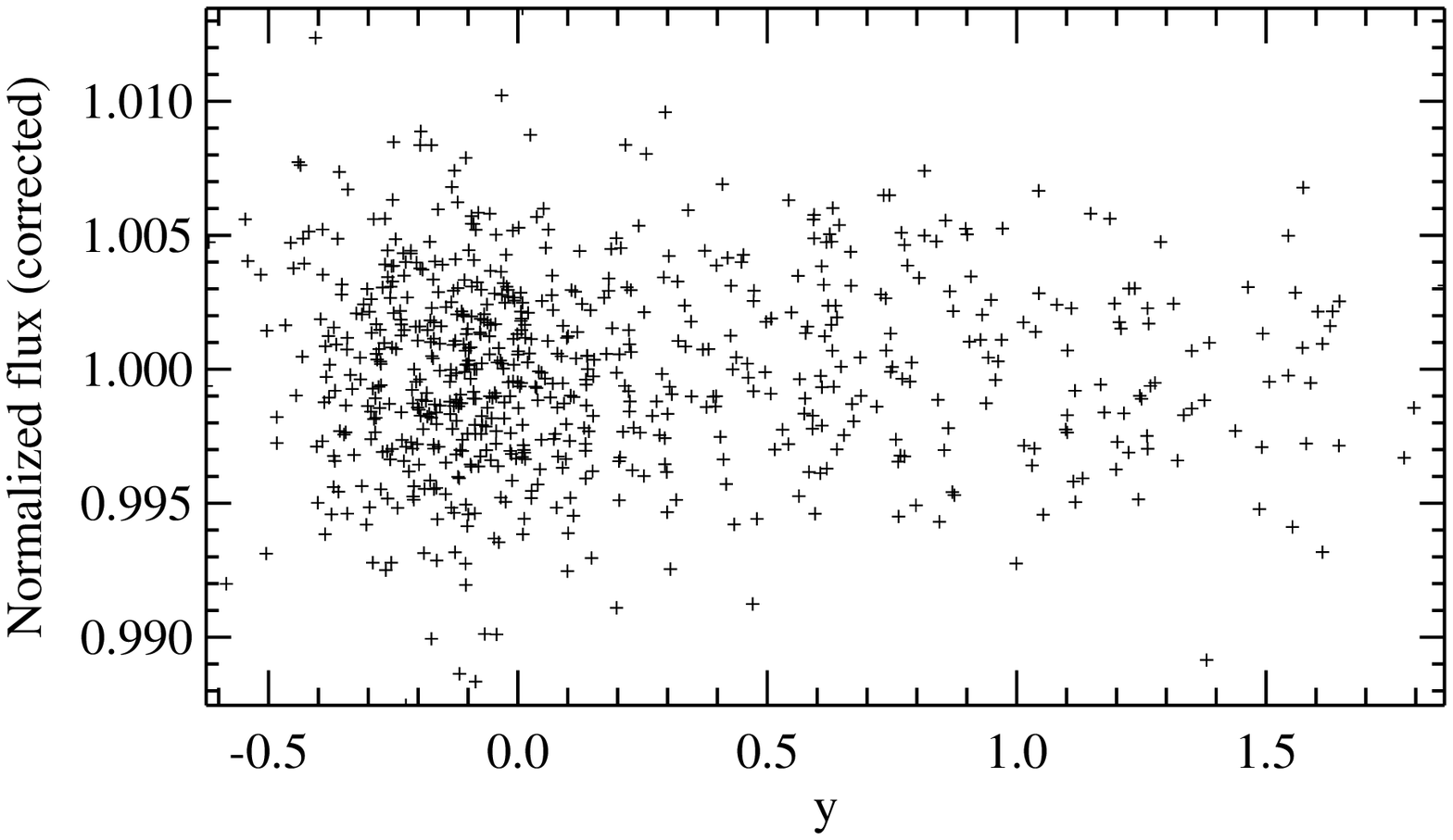}
\includegraphics[width=0.24\textwidth]{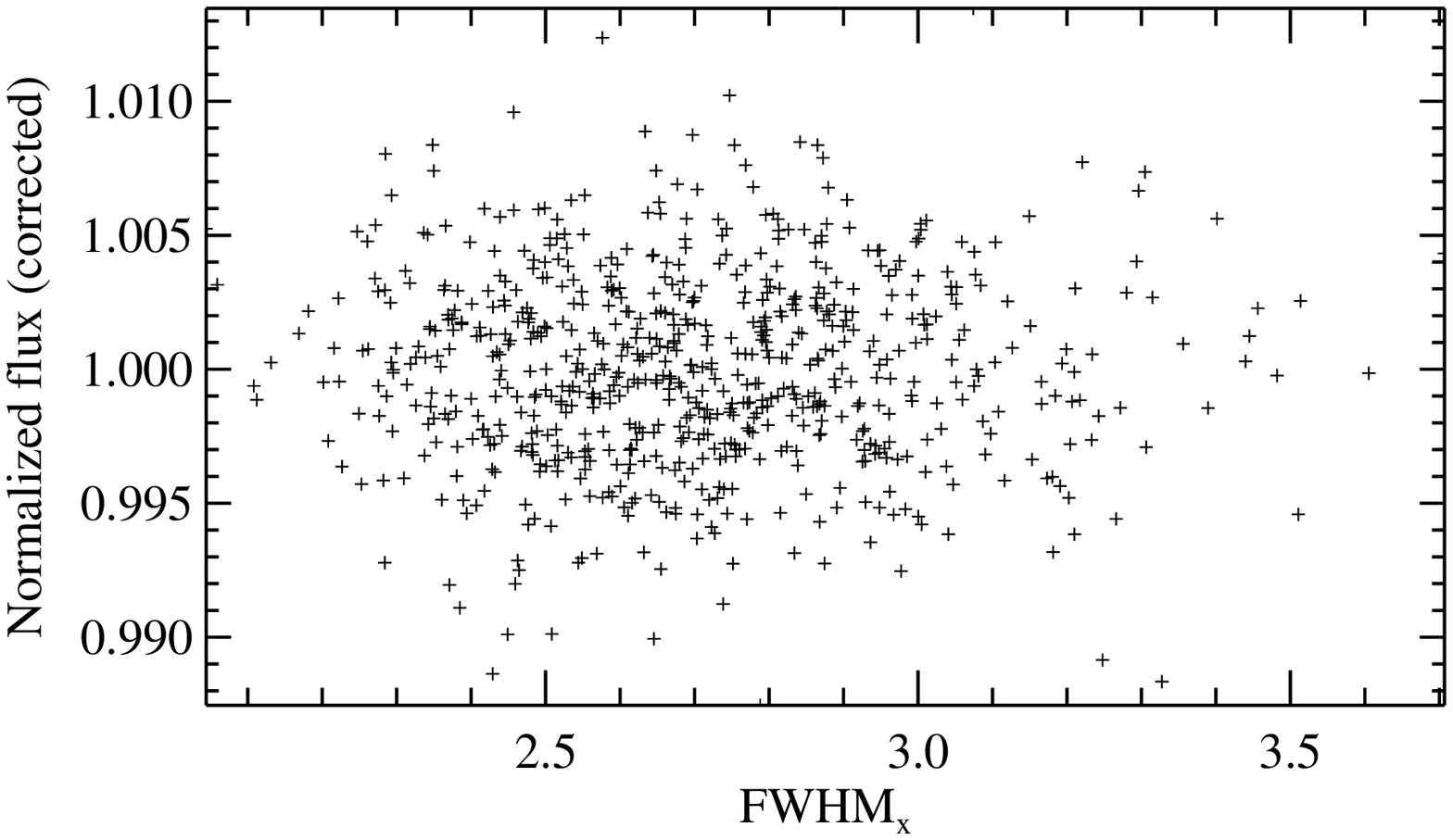}
\includegraphics[width=0.24\textwidth]{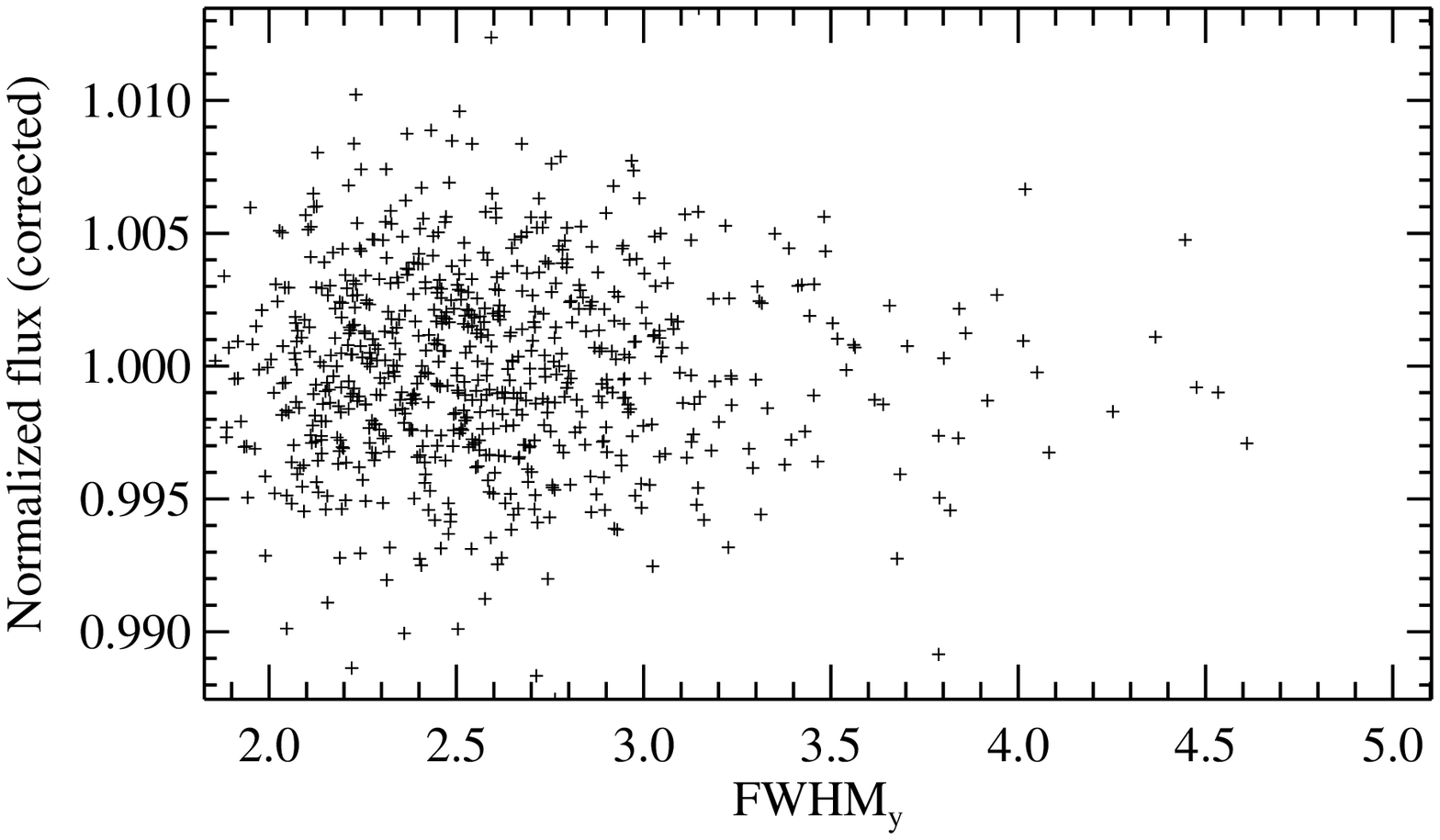}
\caption{Correlation between instrumental parameters (i.e. variables in the baseline function) and the normalized flux for the $K$ band. The {\it first row} shows the flux before baseline correction, while the {\it second row} shows the flux after baseline correction as a comparison. In these plots, $x$ and $y$ refer to the relative positions, and FWHM$_x$ and FHWM$_y$ refer to the full-width at half maximum of the marginalized PSF, both in pixels. }
\label{fig:app_fig3}
\end{center}
\end{figure*}

\begin{figure*}
\begin{center}
\includegraphics[width=0.24\textwidth]{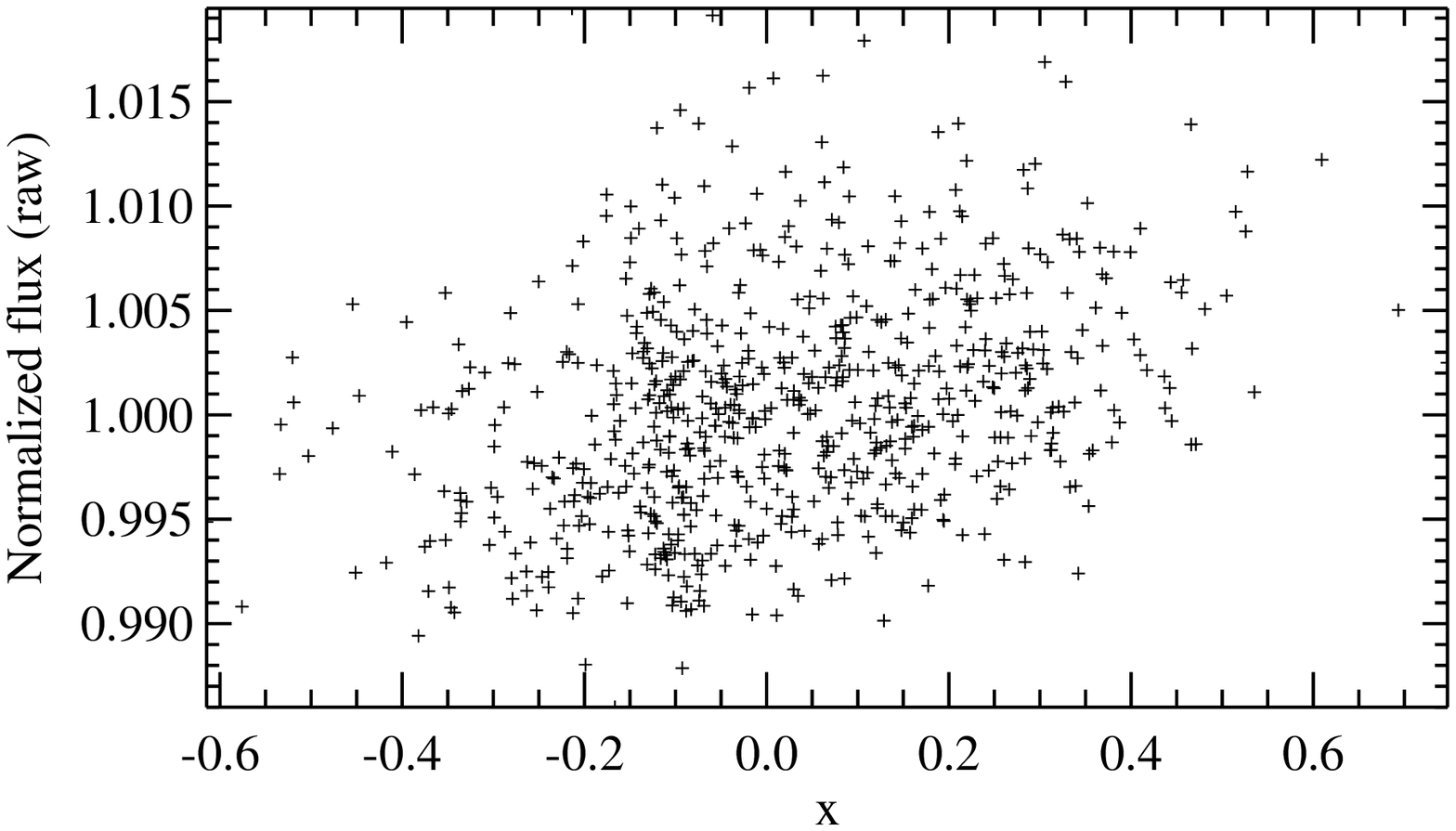}
\includegraphics[width=0.24\textwidth]{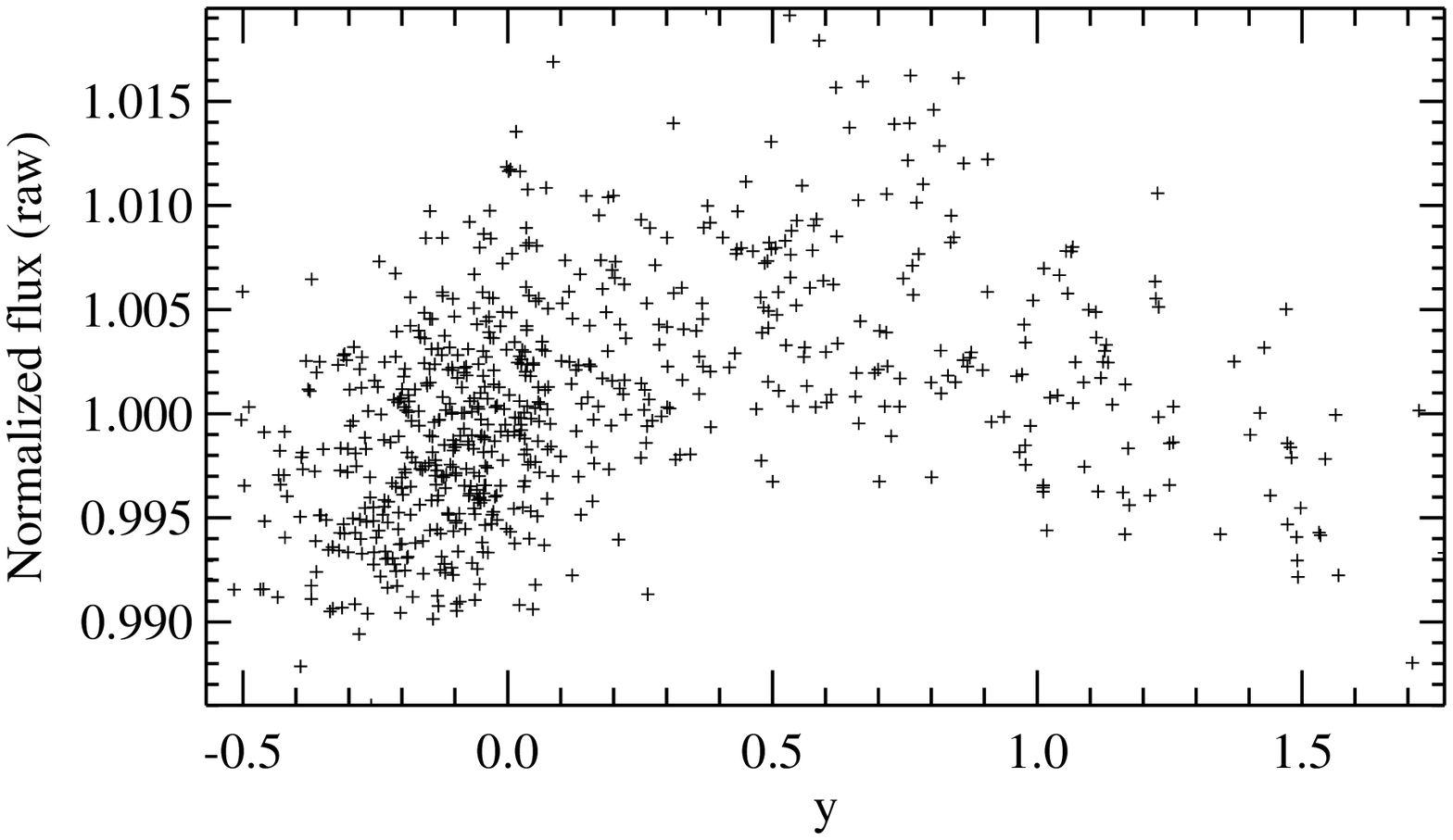}
\includegraphics[width=0.24\textwidth]{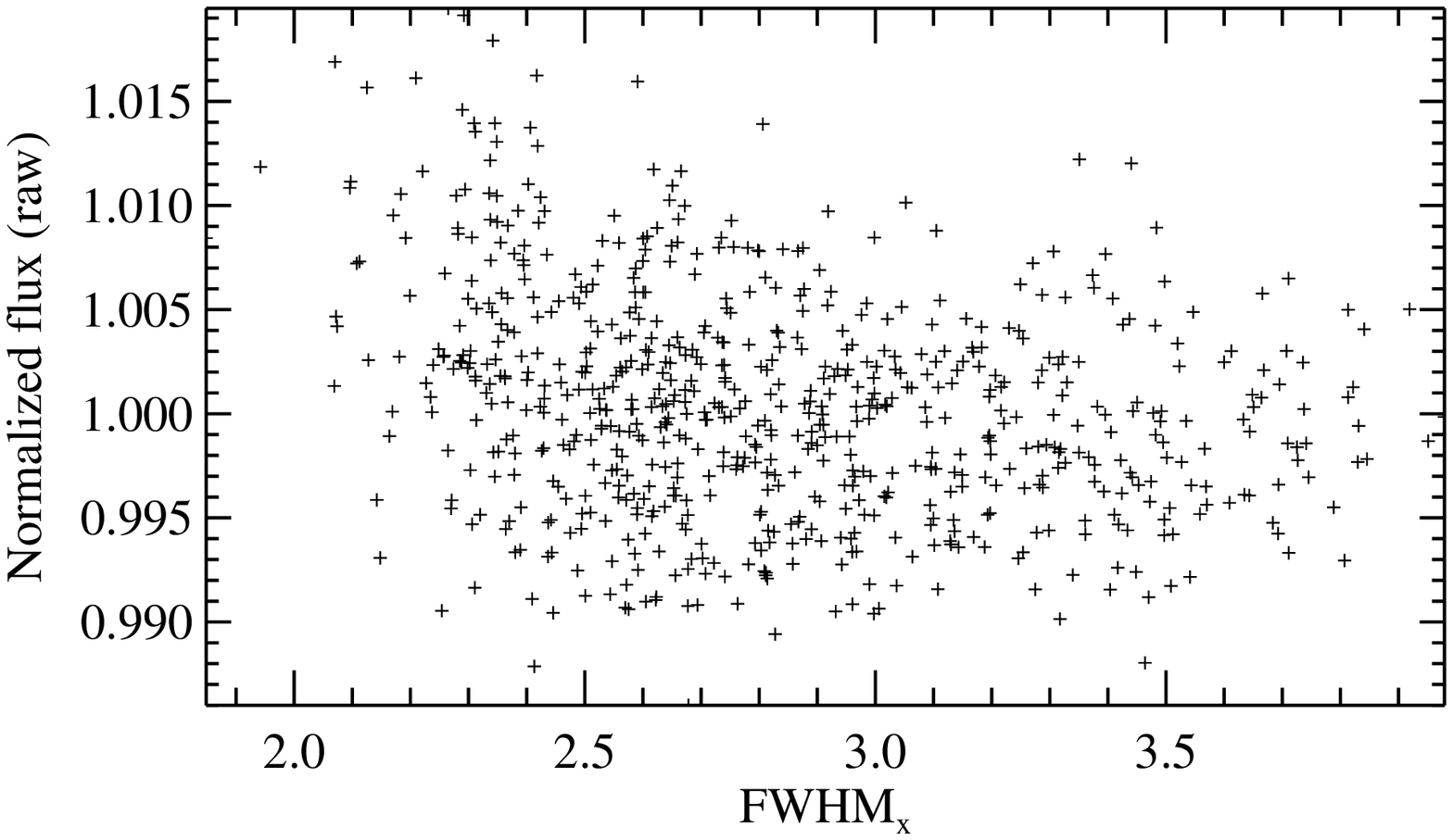}
\includegraphics[width=0.24\textwidth]{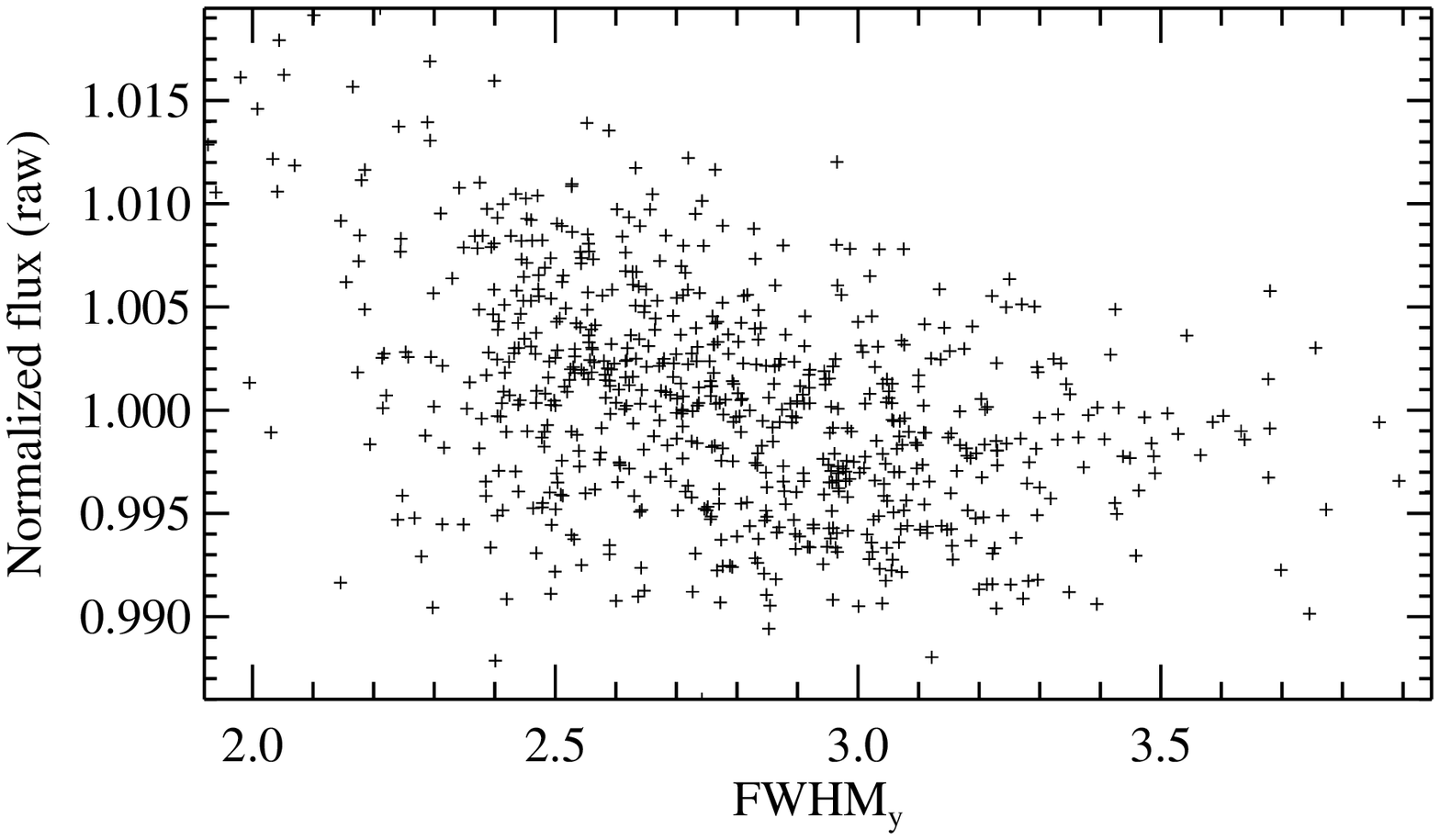}\\
\includegraphics[width=0.24\textwidth]{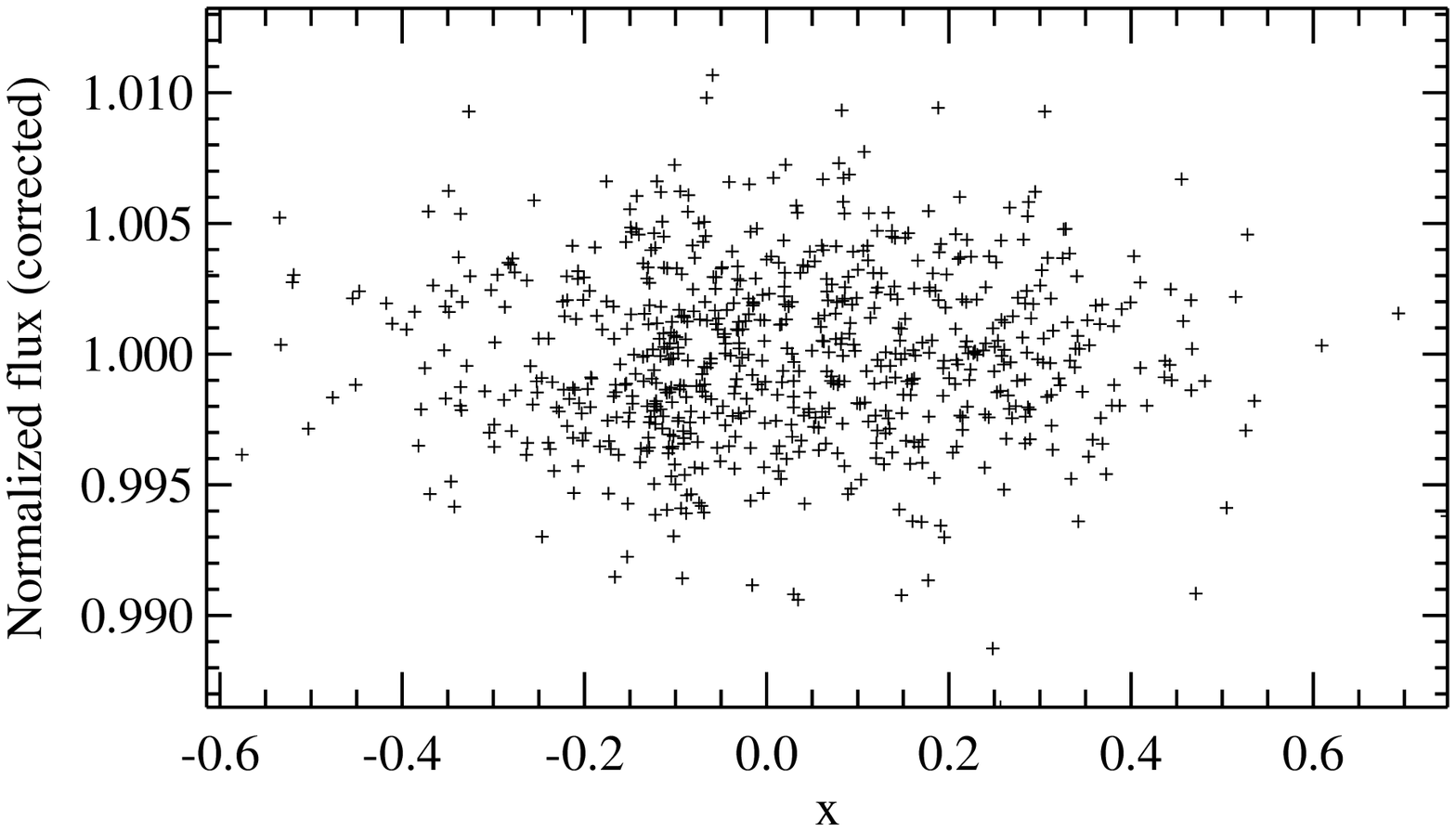}
\includegraphics[width=0.24\textwidth]{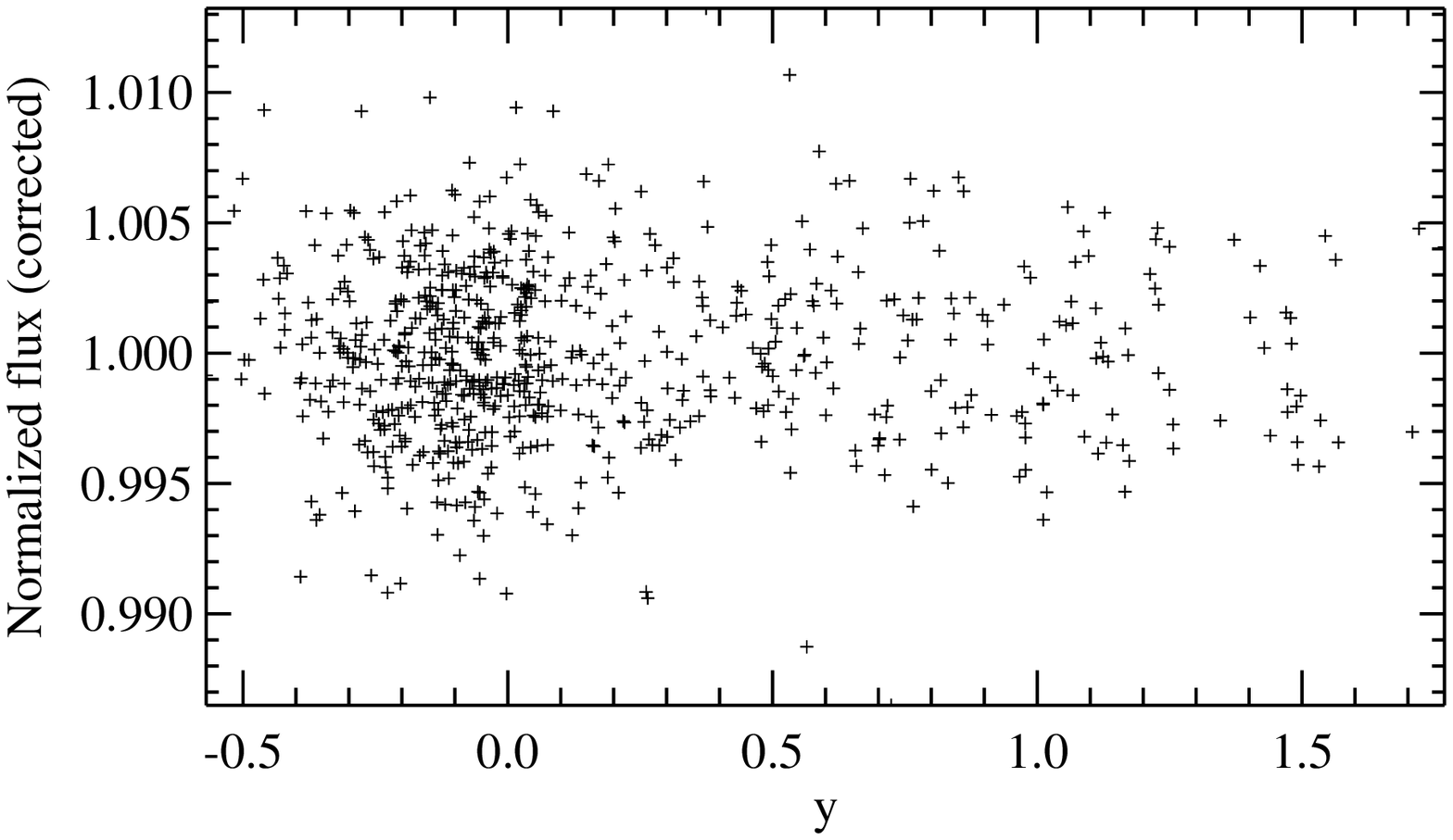}
\includegraphics[width=0.24\textwidth]{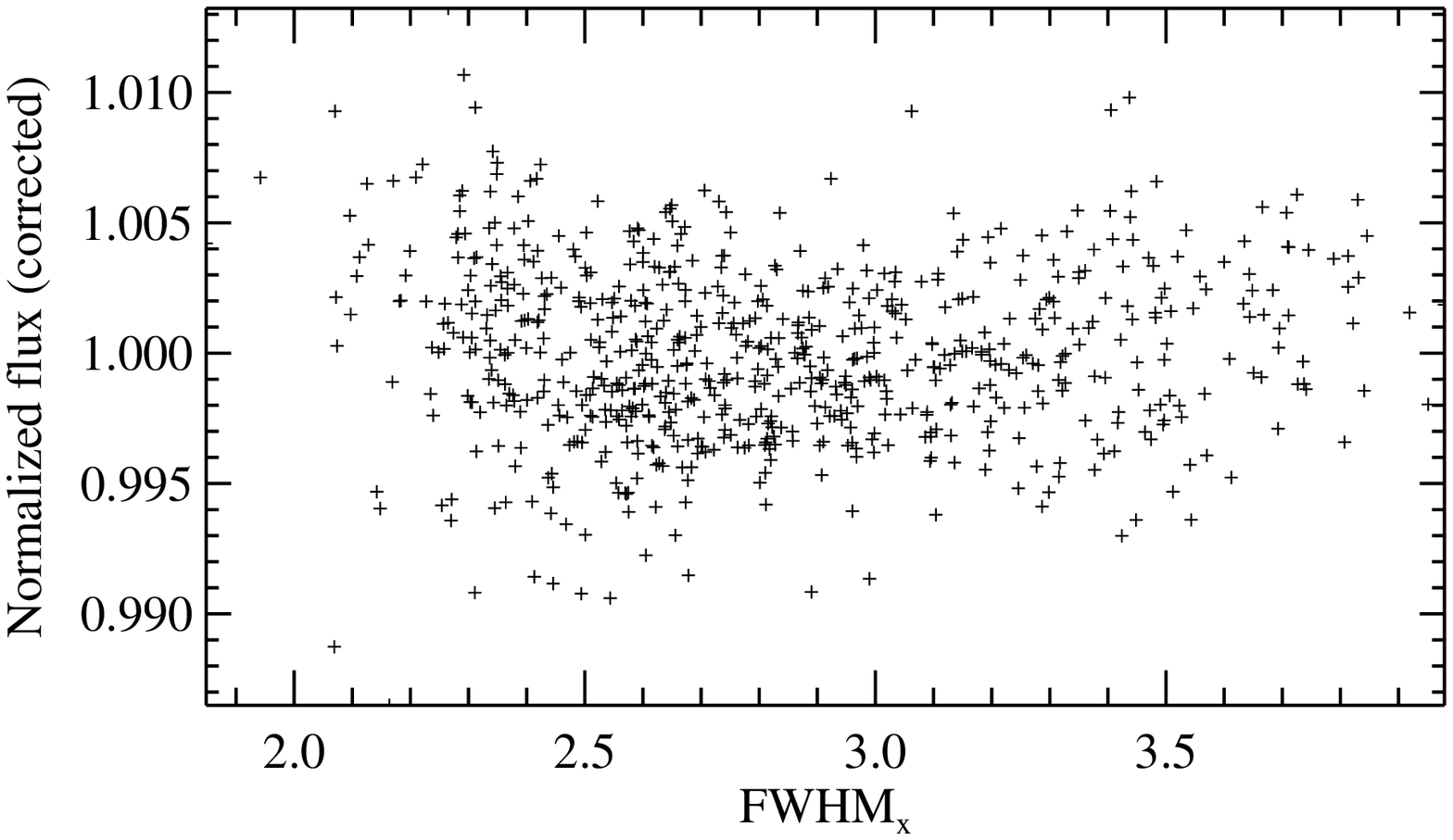}
\includegraphics[width=0.24\textwidth]{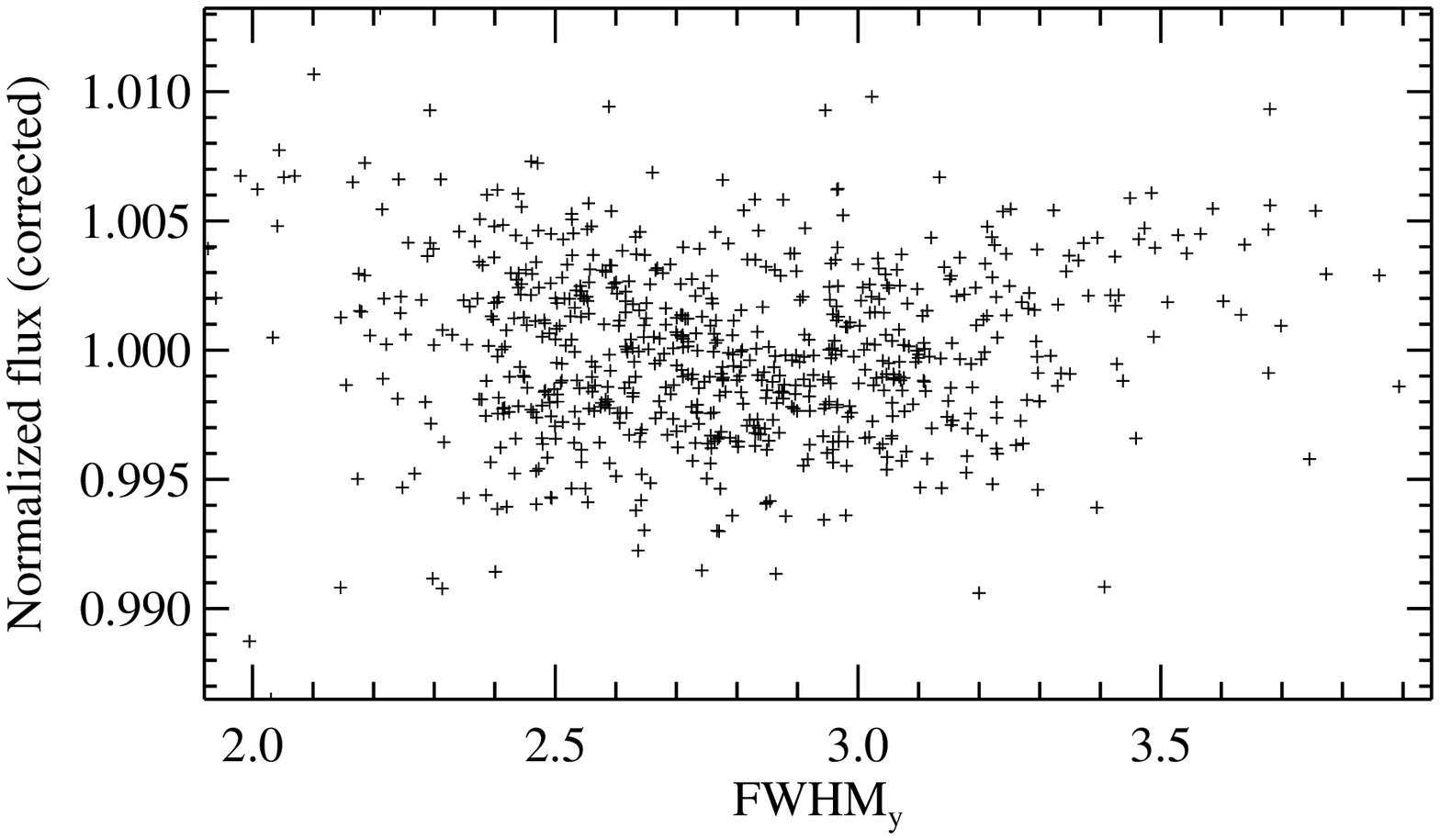}
\caption{Correlation between instrumental parameters and the normalized flux for the $J$ band. The figure is displayed in the same manner as Fig.~\ref{fig:app_fig3}.}
\label{fig:app_fig4}
\end{center}
\end{figure*}

  \end{appendix}

\end{document}